\documentclass{article}

\PassOptionsToPackage{authoryear}{natbib}

\usepackage[preprint]{neurips_2021}

\usepackage[utf8]{inputenc} 
\usepackage[T1]{fontenc}    
\usepackage{hyperref}       
\usepackage{url}            
\usepackage{booktabs}       
\usepackage{amsfonts}       
\usepackage{nicefrac}       
\usepackage{microtype}      
\usepackage{xcolor}         

\usepackage{graphicx}
\usepackage{bbold}
\usepackage{bm}
\usepackage{amsmath,amscd,amssymb}
\usepackage[linesnumbered,ruled,vlined]{algorithm2e}
\usepackage{xcolor,colortbl}
\usepackage{tabu}

\DeclareMathAlphabet{\mathscr}{U}{mathc}{m}{it}

\newcommand{\EE}{\mathbb{E}}
\newcommand{\DD}{\mathcal{D}}
\newcommand{\II}{\mathcal{I}}

\newcommand{\NN}{\mathcal{N}}
\newcommand{\XX}{\mathcal{X}}

\newcommand{\kk}{\mathscr{k}}
\newcommand{\ik}{{\II_\kk}}

\newcommand{\ZZt}{\tilde{\mathcal{Z}}}
\newcommand{\XXt}{\tilde{\mathcal{X}}}
\newcommand{\inv}{^{-1}}
\newcommand{\invT}{^{- {\rm T}}}
\newcommand{\TT}{^{\rm T}}
\newcommand{\ZZ}{\mathcal{Z}}

\newcommand{\OO}{\mathcal{O}}
\newcommand{\RR}{\mathbb{R}}
\newcommand{\ZZZ}{\mathbb{Z}}

\newcommand{\bhat}{\hat{\beta}}
\newcommand{\psihat}{\hat{\psi}}

\newcommand{\id}{\mathbb{1}}
\newcommand{\gmi}{\gamma_{-i}}
\newcommand{\gmj}{\gamma_{-j}}
\newcommand{\olya}{\`olya}

\newcommand{\omhat}{\hat{\omega}}
\newcommand{\Tburn}{T_{\rm burn}}
\newcommand{\TC}{C}
\newcommand{\gone}{{\gamma + 1}}
\newcommand{\gabsone}{{|\gamma| + 1}}
\newcommand{\gabs}{{|\gamma|}}

\title{Fast Bayesian Variable Selection in Binomial and Negative Binomial Regression}

\author{%
  Martin Jankowiak \\
  Broad Institute\\
  Cambridge, Massachusetts, USA\\
  \texttt{mjankowi@broadinstitute.org}
}%

\begin{document}

\maketitle

\begin{abstract}
Bayesian variable selection is a powerful tool for data analysis, as
it offers a principled method for variable selection that accounts for prior information and uncertainty. 
However, wider adoption of Bayesian variable selection has been hampered by computational challenges, 
especially in difficult regimes with a large number of covariates or non-conjugate likelihoods.
Generalized linear models for count data, which are prevalent in biology, ecology, economics, and beyond, represent an important special case.
Here we introduce an efficient MCMC scheme for variable selection in binomial and negative binomial regression
that exploits Tempered Gibbs Sampling \citep{zanella2019scalable} and that includes logistic regression
as a special case. In experiments we demonstrate the effectiveness of our approach, including on cancer
    data with seventeen thousand covariates.
\end{abstract}

\section{Introduction}
\label{sec:intro}

Generalized linear models are ubiquitous throughout applied statistics and data analysis \citep{mccullagh2019generalized}.
One reason for their popularity is their interpretability:
they introduce explicit parameters that encode how the observed response depends
on each particular covariate. 
In the scientific setting this interpretability is of central importance. 
Indeed model fit is often a secondary concern, and the primary goal is to identify
a \emph{parsimonious} explanation of the observed data. 
This is naturally viewed as a model selection problem, in particular one
in which the model space is defined as a nested set of models, with
distinct models including distinct sets of covariates.

The Bayesian formulation of this approach, known as Bayesian variable selection 
in the literature, 
offers a powerful set of techniques for realizing Occam's razor in this 
setting \citep{george1993variable, george1997approaches, chipman2001practical, dellaportas2002bayesian, o2009review}.
Despite the intuitive appeal of this approach, computing or otherwise approximating the
resulting posterior distribution can be computationally challenging. 
A principal reason for this is the astronomical size of the model space that results
whenever there are more than a few dozen covariates.
Indeed for $P$ covariates the total number of distinct models, namely $2^P$, 
exceeds the estimated number of atoms in the known universe ($\sim\!10^{80}$) for $P \gtrsim 266$.
In addition for most models of interest non-conjugate likelihoods make it infeasible to integrate
out real-valued model parameters, resulting in a challenging high-dimensional 
inference problem defined on a transdimensional mixed discrete/continuous latent space. 

In this work we develop efficient MCMC methods for Bayesian variable selection
for two generalized linear models that are commonly used
for analyzing count data: i) binomial regression (and its special 
case logistic regression); and ii) negative binomial regression.
To enable efficient inference we proceed in two steps.
First, we utilize P\olya-Gamma auxiliary variables to establish conjugacy \citep{polson2013bayesian}.
Second, to enable targeted exploration of the high-dimensional model space
we adapt the Tempered Gibbs Sampler (TGS) introduced by \citet{zanella2019scalable} to our setting.
As we will see, this approach ensures that computational resources are strategically 
utilized to explore high probability regions of model space.

\section{Bayesian variable selection in binomial regression}
\label{sec:model}

For simplicity we focus on the binomial regression case, leaving a discussion of the
negative binomial case to Sec.~\ref{app:nb} in the supplemental materials.
Let $X \in \RR^{N\times P}$, $\TC \in \ZZZ^{N}_{> 0}$, and $Y \in \ZZZ^{N}_{\ge 0}$ with $Y \le \TC$ . 
We consider the following space of generalized linear models: 
\begin{align}
\label{eqn:modeldefn}
&{\rm [inclusion \; variables]} \qquad &&\gamma_i \sim {\rm Bernoulli}(h) &&& {\rm for} \;\; i=1,...,P  \\
&{\rm [bias \; term]} \qquad &&\beta_0 \sim \NN(0, \tau_{\rm bias} \inv)  \nonumber\\
&{\rm [coefficients]} \qquad &&\beta_\gamma  \sim \NN(0, \tau \inv \id_\gabs)  \nonumber\\
&{\rm [success \; logits]} \qquad  &&\psi_n \equiv \beta_0 + \beta_\gamma  \cdot X_{n\gamma} &&& {\rm for}  \;\;  n=1,...,N \nonumber \\
&{\rm [responses]} \qquad  &&Y_n \sim {\rm Binomial}(\TC_n, \sigma(\psi_n)) &&& {\rm for}  \;\;  n=1,...,N \nonumber
\end{align}
Here each $\gamma_i \in \{0, 1\}$ controls whether the $i^{\rm th}$ coefficient $\beta_i$---and therefore
the $i^{\rm th}$ covariate---is included ($\gamma_i=1$) or excluded ($\gamma_i=0$) from the model.
In the following we use $\gamma$ to refer to the full $P$-dimensional vector $(\gamma_1, ..., \gamma_P)$.
The hyperparameter $h \in (0, 1)$ controls the overall level of sparsity; in particular $hP$ is the expected number of 
covariates included a priori.\footnote{It is straightforward to place a prior (e.g.~a Beta prior) on $h$.  See \cite{steel2007effect} and references therein for discussion.}
The bias term is governed by a Normal prior with precision $\tau_{\rm bias} > 0$. 
Likewise the $|\gamma|$ coefficients $\beta_\gamma \in \RR^{|\gamma|}$ are governed by a Normal prior with precision $\tau > 0$. For simplicity we take $\tau = \tau_{\rm bias}$ throughout. 
Here $|\gamma| \in \{0, 1, ..., P \}$ denotes the total number of included covariates. 
Note that we assume that the bias term $\beta_0$ is always included. 
The response $Y_n$ is generated from a Binomial
distribution with total count $\TC_n$ and success probability $\sigma(\psi_n)$, where $\sigma(\cdot)$ denotes the logistic or sigmoid function
$\sigma(x) \equiv \{1 + \exp(-x) \}\inv$. 
This reduces to conventional logistic regression with binary responses if $\TC_n=1$ for all $n$.
In the following we drop the $\gamma$ subscript on $\beta_\gamma$ to simplify the notation.

\section{Inference}
\label{sec:inference}


\subsection{P\olya-Gamma augmentation}
\label{sec:pg}

Introducing P\olya-Gamma auxiliary variables is an example of data augmentation, which is commonly used to establish conjugacy or improve mixing \citep{tanner1987calculation}.
P\olya-Gamma augmentation relies on the identity
\begin{align}
    \frac{(e^\psi)^a}{(1 + e^\psi)^b} = 2^{-b} e^{(a-\tfrac{1}{2}b)\psi} \EE_{{\rm PG}(\omega |  b, 0)} \left[ \exp(-\tfrac{1}{2} \omega \psi^2) \right]
\label{eqn:pgidentity}
\end{align}
noted by \citet{polson2013bayesian}. Here $a, \psi \in \RR$, $b >0$, and ${\rm PG}(\omega |  b, 0)$ is the P\olya-Gamma distribution, which
has support on the positive real axis. Using this identity we can introduce a $N$-dimensional vector of P\olya-Gamma 
variates $\omega$ governed by the prior $\omega_n \sim {\rm PG}(\TC_n, 0)$
and rewrite the Binomial likelihood in Eqn.~\ref{eqn:modeldefn} as follows
\begin{align}
p(Y_n | \TC_n, \sigma(\psi_n)) \propto \sigma(\psi_n)^{Y_n} (1 -  \sigma(\psi_n))^{\TC_n - Y_n} 
=  \frac{(\exp(-\psi_n))^{\TC_n - Y_n}}{(1 + \exp(-\psi_n))^{\TC_n}}  = \frac{(\exp(\psi_n))^{Y_n}}{(1 + \exp(\psi_n))^{\TC_n}}  \nonumber 
\end{align}
so that each Binomial likelihood term in Eqn.~\ref{eqn:modeldefn} is now replaced with a factor 
\begin{equation}
\label{eqn:pgfactor}
\exp(\kappa_n \psi_n -\tfrac{1}{2} \omega_n \psi_n^2)          \qquad {\rm with} \qquad          \kappa_n \equiv Y_n - \tfrac{1}{2} \TC_n
\end{equation}
This augmentation leaves the marginal distribution w.r.t.~$(\gamma, \beta)$ unchanged.
Crucially each factor in Eqn.~\ref{eqn:pgfactor} is Gaussian w.r.t.~$\beta$, with
the consequence that P\olya-Gamma augmentation establishes conjugacy.

\subsection{Binary variables and Metropolized-Gibbs}
\label{sec:mg}

The augmented model in Sec.~\ref{sec:pg} readily admits a conventional Gibbs sampling scheme.\footnote{In particular one that
alternates between $\omega$ updates and $(\gamma_i, \beta)$ updates.}
However, the mixing time of the resulting sampler is notoriously slow in high dimensions:
the sampler is \emph{sticky}. For example, consider the scenario in which the two covariates corresponding to $i=1$ and $i=2$ are highly correlated and
each on its own is sufficient for explaining the observed data. In this scenario the posterior concentrates on models with
$\gamma = (1, 0, 0, 0, ...)$ and $\gamma = (0, 1, 0, 0, ...)$. Unfortunately, single-site Gibbs updates w.r.t.~$\gamma_i$ will move between the two modes very
infrequently, since they are separated by low probability models like $\gamma = (0, 0, 0, 0, ...)$ and $\gamma = (1, 1, 0, 0, ...)$.

To motivate the Tempered Gibbs Sampling (TGS) strategy that we use to deal with this stickiness, we consider a single latent binary variable $x$   
governed by the probability distribution $p(x) = {\rm Bernoulli}(q)$. A Gibbs sampler for this distribution simply samples $x \sim p$ in
each iteration of the Markov chain. An alternative strategy is to employ a so-called Metropolized-Gibbs move w.r.t.~$x$ \citep{liu1996peskun}. 
For binary $x$ this results in a proposal distribution that is deterministic in the sense that it always proposes a flip: $0 \to 1$ or $1 \to 0$.
The corresponding Metropolis-Hastings (MH) acceptance probability for a proposed move $x \to x^\prime$ is given by
\begin{align}
\alpha(x \! \to \! x^\prime) = 
    \begin{cases}
      \min(1, \tfrac{q}{1-q}) & {\rm if} \; x = 0\\
     \min(1, \tfrac{1- q}{q}) & {\rm if} \; x = 1
    \end{cases} 
    \label{eqn:accept}
\end{align}
As is well known, this update rule is more statistically efficient than the corresponding Gibbs move \citep{liu1996peskun}. 
For our purposes, however, what is particularly interesting is the special case where $q=\tfrac{1}{2}$. In this
case the acceptance probability in Eqn.~\ref{eqn:accept} is identically equal to one. Consequently
the Metropolized-Gibbs chain is deterministic: 
\begin{align}
... \to 0\to 1 \to 0 \to 1 \to 0 \to 1 \to ...
\end{align}
Indeed this Markov chain can be described as \emph{maximally non-sticky}. 

\subsection{A tempered inference scheme}
\label{sec:tempinf}

We now describe how we adapt the Tempered Gibbs Sampling (TGS) strategy in \citet{zanella2019scalable} to our setting.
The basic idea of TGS in this context is to introduce a \emph{tempered} auxiliary target distribution 
that leverages the non-sticky dynamics described in Sec.~\ref{sec:mg}.
Importantly, this tempered distribution needs to be sufficiently close to the PG-augmented distribution described in Sec.~\ref{sec:pg} so
that we can use importance sampling to obtain samples from the PG-augmented distribution while avoiding high-variance importance weights.

In more detail we proceed as follows. The augmented target distribution in Sec.~\ref{sec:pg} is given by
\begin{align}
p(Y | \beta, \gamma, \omega, X, \TC)  p(\beta)  p(\gamma)p(\omega|\TC)  &= Z p(\beta,\gamma, \omega | \DD) 
\label{eqn:augtargetbeta}
\end{align}
where we define $\DD \equiv \{X, Y, \TC\}$ so that $p(\beta, \gamma, \omega | \DD)$ is the corresponding posterior
and $Z$ is the (intractable) normalization constant. Since the dimension of $\beta$ depends on $\gamma$,
it is advantageous to marginalize out $\beta$ to obtain
\begin{align}
p(Y | \gamma, \omega, X, \TC) p(\gamma) p(\omega|\TC) = Z p(\gamma, \omega | \DD) 
\label{eqn:augtarget}
\end{align}
Thanks to P\olya-Gamma augmentation we can compute $p(Y | \gamma, \omega,  X, \TC)$ in closed form.
The next step is to induce \emph{coordinatewise tempering}  in Eqn.~\ref{eqn:augtarget}.
To do this we introduce an auxiliary variable $i \in \{0, 1, 2, ..., P\}$ that controls which variables, if any, are tempered.
 In particular we define the following tempered (unnormalized) target distribution:
\begin{align}
\label{eqn:fulltarget}
f(\gamma, \omega, i) 
&\equiv Z p(\gamma, \omega | \DD)  \left \{ \delta_{i0} \xi
                   + \frac{1}{P} \sum_{j=1}^P \delta_{ij} \eta(\gmj, \omega) \frac{U(\gamma_j)}{ p(\gamma_j | \gmj, \omega, \DD)} \right\} \\                   
&= \delta_{i0} \xi Z p(\gamma, \omega | \DD) 
                   + \frac{Z}{P} \sum_{j=1}^P \delta_{ij} \eta(\gmj, \omega)  U(\gamma_j) p( \gmj, \omega | \DD)                  \nonumber
\end{align}
Here $\xi > 0$ is a hyperparameter whose choice we discuss below and  $\delta_{ij}$ is the 
Kronecker delta.\footnote{That is $\delta_{ij}=1$ if $i=j$ and $0$ otherwise.}
Furthermore $U(\cdot)$ is the uniform distribution on $\{0, 1\}$ and $\gmj$ denotes all components of $\gamma$ apart
from $\gamma_j$. Finally $\eta(\gmj, \omega)$ is an additional weighting factor to be defined later; for now the reader can
suppose that $\eta(\gmj, \omega)=1$.

We note three important features of Eqn.~\ref{eqn:fulltarget}.
First, since $i$ is included explicitly as a variable in Eqn.~\ref{eqn:fulltarget} we can use $i$ to guide our sampling strategy without
violating the conditions of a Markov chain.
Second, by construction when $i>0$ the posterior conditional w.r.t.~$\gamma_i$ is the uniform distribution $U(\gamma_i)$.
Consequently for $i>0$ Metropolized-Gibbs updates w.r.t.~$\gamma_i$ will result in the desired non-sticky dynamics described in the previous section.
Third, as we discuss in more detail in Sec.~\ref{app:mll}, the posterior conditional $p(\gamma_i | \gmi, \omega, \DD)$ in Eqn.~\ref{eqn:fulltarget}
can be computed in closed form thanks to P\olya-Gamma augmentation. This is important because computing $p(\gamma_i | \gmi, \omega, \DD)$ 
is necessary for importance weighting as well as computing Rao-Blackwellized posterior inclusion probabilities.

We proceed to construct a sampling scheme for the target distribution Eqn.~\ref{eqn:fulltarget} that utilizes Gibbs updates w.r.t.~$i$,
Metropolized-Gibbs updates w.r.t.~$\gamma_i$, and Metropolis-Hastings updates w.r.t.~$\omega$.

\paragraph{$i$-updates}

If we marginalize $i$ from Eqn.~\ref{eqn:fulltarget}  we obtain
\begin{align}
f(\gamma, \omega) = Z p(\gamma, \omega | \DD)  \left \{ \xi
                   + \frac{1}{P} \sum_{j=1}^P \eta(\gmj, \omega) \frac{U(\gamma_j)}{ p(\gamma_j | \gmj, \omega, \DD)} \right\}  = Z p(\gamma, \omega | \DD)   \phi(\gamma, \omega)
\label{eqn:fphi}
\end{align}
where we define 
\begin{align}
 \phi(\gamma, \omega) \equiv \xi + \frac{1}{P} \sum_{i=1}^P  \frac{\tfrac{1}{2} \eta(\gmi, \omega)}{ p(\gamma_i | \gmi, \omega, \DD)}
\label{eqn:phidefn}
\end{align}
As is clear from Eqn.~\ref{eqn:fphi}, $\phi(\gamma, \omega)\inv$ is the importance weight that is used to obtain samples 
from the non-tempered target Eqn.~\ref{eqn:augtarget}. 
Additionally these equations imply that we can do Gibbs updates w.r.t.~$i$ using the (normalized) distribution
\begin{align}
\label{eqn:iupdate}
f(i | \gamma, \omega)  =  \frac{1}{ \phi(\omega, \gamma) } \left( \delta_{i0} \xi + \frac{1}{P} \sum_{j=1}^P \delta_{ij} \frac{\tfrac{1}{2}  \eta(\gmj|\omega)}{ p(\gamma_j | \gmj, \omega, \DD)} \right)
\end{align}

To better understand the behavior of the auxiliary variable $i$, we compute the marginal 
distribution w.r.t.~$i$ for the special case $\eta(\cdot)=1$:
\begin{align}
f(i) \propto \delta_{i0} \xi   + \frac{1 }{P} \sum_{j=1}^P \delta_{ij} 
\label{eqn:fmarg}
\end{align}
We can read off several conclusions from Eqn.~\ref{eqn:fmarg}.
First, $\xi$ controls how frequently we visit the $i=0$ state. Second, if $\eta(\cdot)=1$ then the
states with $i > 0$ (each of which corresponds to a particular covariate in X) are visited equally often. Later we
discuss how we can choose $\eta(\cdot)$ to preferentially visit certain states and thus preferentially update $\gamma_i$ for particular covariates.

\paragraph{$\gamma$-updates}

Whenever $i > 0$ we update $\gamma_i$.  As discussed above, for $i>0$ the posterior conditional w.r.t.~$\gamma_i$ is the uniform distribution $U(\gamma_i)$.
We employ Metropolized-Gibbs updates w.r.t.~$\gamma_i$, resulting in deterministic flips that are accepted with probability one:
$\gamma_i \to 1 - \gamma_i$.

\paragraph{$\omega$-updates}

Whenever $i = 0$ we update $\omega$.  To do so we use a simple proposal distribution that can be computed in closed form.
Importantly $f(\gamma, \omega, i=0)$ is not tempered so we can rely on the conjugate structure that is made manifest when we condition
on an explicit value of $\beta$.
In more detail, we first compute the \emph{mean} of the conditional posterior distribution $p(\beta | \gamma, \omega, \DD)$ of Eqn.~\ref{eqn:augtargetbeta}:
\begin{align}
    \bhat(\gamma, \omega) \equiv \EE_{p(\beta | \gamma, \omega, \DD)} \left[ \beta \right]
\end{align}
Using this (deterministic) value we then form the conditional posterior distribution $p(\omega^\prime | \gamma, \bhat, \DD)$, which is a P\olya-Gamma distribution
whose parameters are readily computed.
We then sample a proposal $\omega^\prime \sim p(\cdot | \gamma, \bhat, \DD)$ and 
compute the corresponding MH acceptance probability $\alpha(\omega \! \to \! \omega^\prime | \gamma)$. The proposal is then accepted
with probability $\alpha(\omega  \! \to \! \omega^\prime | \gamma)$; otherwise it is rejected. The acceptance probability 
can be computed in closed form and is given by
\begin{align}
\label{eqn:alphaomega}
\!\!\!    \alpha(\omega  \! \to \! \omega^\prime | \gamma) &= \min\!\left(\!1, 
\frac{p(Y | \gamma, \omega^\prime, X, \TC) }{p(Y | \gamma, \omega, X, \TC)}
\frac{p(Y | \gamma, \omega, \bhat(\gamma, \omega^\prime), X, \TC) }{p(Y | \gamma, \omega^\prime, \bhat(\gamma, \omega), X, \TC)}
    \frac{p(Y | \gamma,  \bhat(\gamma, \omega), X, \TC) }{p(Y | \gamma, \bhat(\gamma, \omega^\prime), X, \TC)} \right)
\end{align}
Each of the terms in Eqn.~\ref{eqn:alphaomega} can be readily computed; conveniently there is no need to compute
the P\olya-Gamma density, which can be challenging in some regimes. See Sec.~\ref{app:omega} for more details.

We note that the proposal distribution $p(\omega^\prime | \gamma, \bhat(\gamma, \omega), \DD)$ can be thought of as an approximation to the posterior conditional
$p(\omega^\prime | \gamma, \DD) = \int \! d\beta \, p(\omega^\prime | \gamma, \beta, \DD) p(\beta | \gamma, \DD)$ that would be used in a Gibbs update. 
Since this latter density is intractable, we instead opt for this tractable option.
One might worry that the resulting acceptance probability
would be very low, since $\omega$ is $N$-dimensional and $N$ can be large. 
However, $p(\omega^\prime | \gamma, \beta, \DD)$ only depends
on $\beta$ through the scalars $\psi_n = \beta_\gamma \cdot X_{n\gamma}$; the induced posterior over $\psi_n$ is typically somewhat narrow, since 
the $\psi_n$ are pinned down by the observed data, and consequently  $p(\omega^\prime | \gamma, \bhat, \DD)$ is a reasonably good approximation to the
exact posterior conditional. 
In practice we observe high mean acceptance probabilities $\sim50\% - 95\%$ for all the experiments in this work,\footnote{For example for the experiment in Sec.~\ref{sec:runtime}, in which
$N$ varies between $100$ and $4000$ and $P$ varies between $134$ and $69092$, the
average acceptance probability ranges between $49\%$ and $89\%$.}
even for $N \gg 10^2$, although we note that we would expect problematically small acceptance probabilities in sufficiently extreme regimes, for example if both $N$ and $\TC$ are large.
 
\paragraph{Weighting factor $\eta$}

To finish specifying our inference scheme we need to choose the weighting factor $\eta(\gmj, \omega)$ in Eqn.~\ref{eqn:fulltarget}.
As discussed after Eqn.~\ref{eqn:fmarg}, if we choose $\eta(\cdot)=1$ the $P$ states corresponding to $i > 0$ will be explored with
equal frequency. For covariates $i$ that have very small posterior inclusion probabilities (PIPs), i.e.~$p(\gamma_i = 1 | \DD) \ll 1$,
this non-preferential exploration strategy will tend to waste computation on exploring parts of $\gamma$ space with low posterior probability. 
By choosing $\eta(\cdot)$ accordingly we can steer our computational resources towards high probability regions of the posterior.
Similarly to \citet{zanella2019scalable} we let 
\begin{align}
\eta(\gmj, \omega) = p(\gamma_j = 1 | \gmj, \omega) + \tfrac{\epsilon}{P}
\label{eqn:etadefn}
\end{align}
Here $p(\gamma_j = 1 | \gmj, \omega) $ is a conditional PIP, and $\epsilon$ is a hyperparameter that trades off between
 exploitation ($\epsilon \to 0$) and exploration ($\epsilon \to \infty$).

For the full algorithm see Algorithm~\ref{algo}. Following  \citet{zanella2019scalable} we call the algorithm with trivial $\eta(\cdot)$
TGS and the algorithm with non-trivial $\eta(\cdot)$ Weighted TGS (wTGS).
Furthermore we call the algorithm without tempering but with $\eta(\cdot)$ as in Eqn.~\ref{eqn:etadefn} wGS. Like (w)TGS this algorithm utilizes Metroplized-Gibbs moves w.r.t.~$\gamma_i$, although in contrast to (w)TGS the $\gamma_i$ update
is \emph{not} deterministic.\footnote{Instead an acceptance probability of the form Eqn.~\ref{eqn:accept} is used.}

\begin{algorithm*}[t!]
\DontPrintSemicolon 
\KwIn{Dataset $\DD = \{X, Y, \TC\}$ with $N$ data points and $P$ covariates;
     total number of MCMC iterations $T$; 
     number of burn-in iterations $\Tburn$;
     hyperparameter $\xi > 0$ (optional)
     }
    \KwOut{Approximate weighted posterior samples $\{\rho_t, \gamma_t, \omega_t \}^T_{t=\Tburn +1}$}
    Let $\gamma_0 = (0, ..., 0)$ and $\omega_0 \sim {\rm PG}(\TC, 0)$. \\
\For{$t =1, ..., T$} {
    Sample $i_t \sim f(\cdot | \gamma_{t-1}, \omega_{t-1})$ using Eqn.~\ref{eqn:iupdate} \\
    If $i_t >0$ let $\gamma_t = {\rm flip}(\gamma_{t-1} | i_t)$ where ${\rm flip}( \gamma | i)$ 
        flips the $i^{\rm th}$ coordinate of $\gamma$: $\gamma_i \to 1 - \gamma_i$. \\
    Otherwise if $i_t =0$ sample $\omega_t^\prime \sim p(\cdot | \gamma_{t-1}, \bhat(\gamma_{t-1}, \omega_{t-1}), \DD)$.
    Set $\omega_t = \omega_t^\prime$ with probability $\alpha(\omega  \! \to \! \omega^\prime | \gamma)$ given in Eqn.~\ref{eqn:alphaomega}.
    Otherwise set $\omega_t = \omega_{t-1}$. \\
    Compute the unnormalized weight $\tilde{\rho}_t = \phi(\gamma_t, \omega_t)^{-1}$ using Eqn.~\ref{eqn:phidefn}. \\
    If $\xi$ is not provided and $t \le \Tburn$ adapt $\xi$ using the scheme described in Sec.~\ref{app:xi}.
}
    Compute the normalized weights $\rho_t = \frac{ \tilde{\rho}_t  } { \sum_{s > \Tburn} \tilde{\rho}_s }$ for 
    $t=\Tburn + 1,...,T$. \\
    \Return{$\{\rho_t, \gamma_t, \omega_t \}_{t=\Tburn + 1}^T$}
    \caption{We outline the main steps in (w)TGS. wTGS uses the weighting function $\eta(\cdot)$ in
             Eqn.~\ref{eqn:etadefn} and TGS uses $\eta(\cdot)=1$. See Sec.~\ref{sec:inference} for details.}
\label{algo}
\end{algorithm*}

\paragraph{Importance weights}

The Markov chain in Algorithm~\ref{algo} targets the auxiliary distribution Eqn.~\ref{eqn:fulltarget}.
To obtain samples from the desired posterior $p(\gamma, \omega | \DD)$ we must reweight each sample $(\gamma, \omega)$
with an importance weight given by $\rho = \phi(\gamma, \omega)^{-1}$; see Eqn.~\ref{eqn:fphi}.
Crucially, for TGS with $\eta(\cdot)=1$ we necessarily have $\rho \le (\xi + \tfrac{1}{2})\inv $.
Likewise for wTGS we necessarily have $\rho \le (\xi + \tfrac{\epsilon}{2P})\inv$. We typically choose a value
for $\xi$ in the range $\xi \sim 1-5$. Consequently the importance weights $\rho$ are bounded from above by a $\OO(1)$ constant
and exhibit only moderate variance. Ultimately this moderate variance can be traced to the coordinatewise tempering scheme used
in TGS, which keeps the tempering to a modest level. Nevertheless as we show in experiments in Sec.~\ref{sec:exp}, 
this modest amount of tempering leads to dramatic improvements in statistical efficiency.

\section{Related work}
\label{sec:related}

Some of the earliest approaches to Bayesian variable selection (BVS) were introduced by \citet{george1993variable, george1997approaches}.
\citet{chipman2001practical} provide an in-depth discussion of BVS for linear regression and CART models.
\citet{zanella2019scalable} introduce Tempered Gibbs Sampling (TGS) and apply it to BVS for linear regresion.
\citet{nikooienejad2016bayesian} and \citet{shin2018scalable} advocate the use of non-local priors in the BVS setting.
\citep{garcia2013sampling} compare non-MCMC search methods for BVS to MCMC-based methods. 
\citet{griffin2021search} introduce an efficient adaptive MCMC method for BVS in linear regression. \citet{wan2021adaptive} extend this approach to logistic regression and accelerated failure time models. We include this approach (ASI) in our empirical validation in Sec.~\ref{sec:exp}.
\citet{tian2019bayesian} introduce an approach to BVS for logistic regression that relies on joint credible regions.
\citet{lamnisos2009transdimensional} develop transdimensional MCMC chains for BVS in probit regression.
\citet{dellaportas2002bayesian} and \citet{o2009review} review various methods for BVS. 
\citet{polson2013bayesian} introduce P\olya-Gamma augmentation and use it to develop efficient Gibbs samplers.

\section{Experiments}
\label{sec:exp}

In this section we validate the performance of Algorithm~\ref{algo} on synthetic and real world data.
We implement all algorithms using PyTorch \citep{paszkepytorch} and leverage
the \texttt{polyagamma} package for sampling from P\olya-Gamma distributions \citep{bleki_2021}.
Besides our main algorithm wTGS, we consider its two simpler variants TGS and wGS. 
In addition we compare to ASI \citep{wan2021adaptive}, an adaptive MCMC
scheme for logistic (and thus binomial) regression that likewise uses P\olya-Gamma augmentation.
In Sec.~\ref{sec:runtime}-\ref{sec:cancer} we consider binomial and logistic regression
and in Sec.~\ref{sec:hospital}-\ref{sec:health} we consider negative binomial regression.

\subsection{Runtime}
\label{sec:runtime}

\begin{figure}[ht]
\centering
    \includegraphics[width=0.7\linewidth]{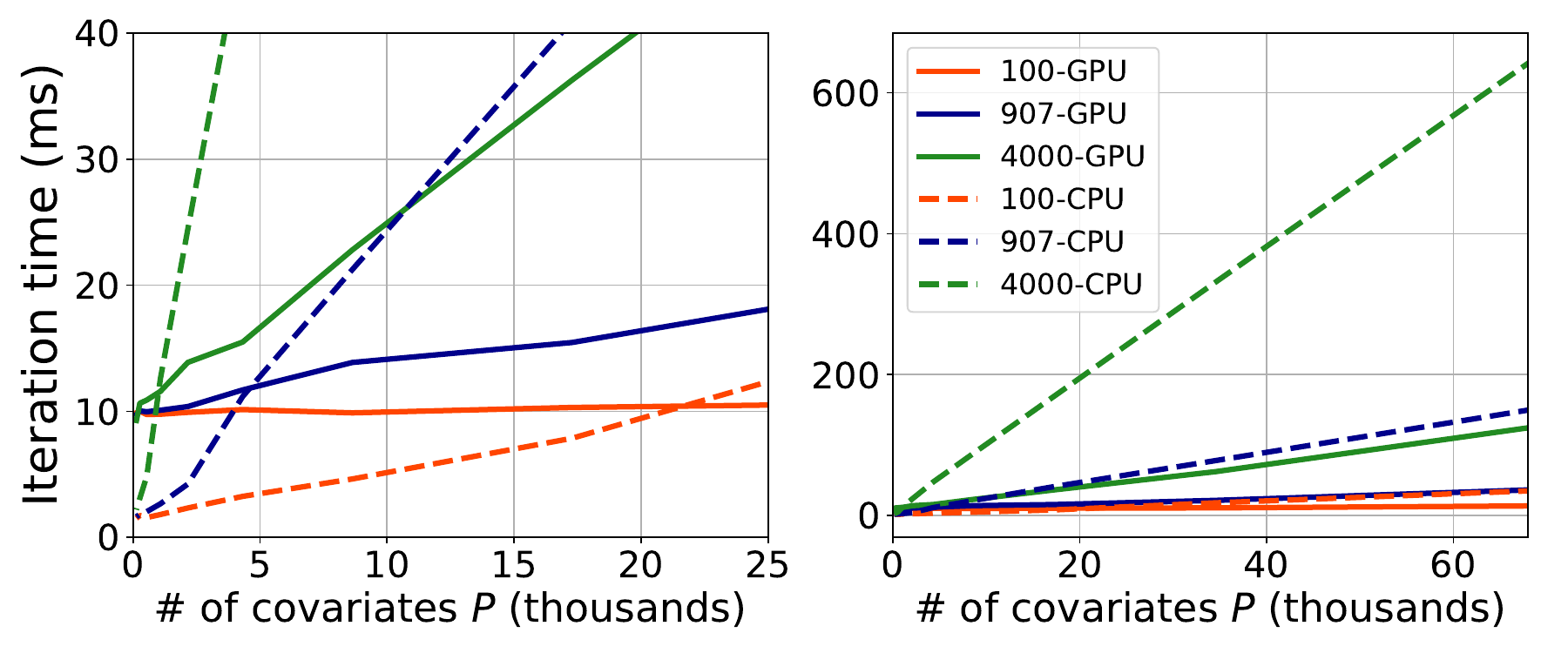}
\caption{We depict MCMC iteration times in milliseconds for wTGS on CPU and GPU as the number of
    covariates $P$ is varied. We also vary the number of data points $N \in \{ 100, 907, 4000 \}$.
    See Sec.~\ref{sec:runtime} for details. 
    Note that the figure on the left is a magnified version of the figure on the right.
    The CPU has 24 cores (Intel Xeon Gold 5220R 2.2GHz) and the GPU is a NVIDIA Tesla K80 GPU.}
\label{fig:runtime}
\end{figure}

In Fig.~\ref{fig:runtime} we depict MCMC iteration times for wTGS for various values
of $N$ and $P$. To make the benchmark realistic we use semi-synthetic data derived from 
the DUSP4 cancer dataset ($N=907$, $P=17273$) used in Sec.~\ref{sec:cancer}.\footnote{In particular
for $N \ne 907$ and $P \ne 17273$ we subsample and/or add noisy data point replicates and/or 
add random covariates as needed.} As discussed in more detail in Sec.~\ref{app:cc} in the supplemental
materials, all four algorithms have similar runtimes, since each is dominated by the $\OO(P)$ cost
of computing $p(\gamma_j = 1 | \gmj, \omega)$ for $j=1,...,P$. As can be seen in Fig.~\ref{fig:runtime}
for any given $N$ the iteration time is lower on CPU for small $P$, but GPU parallelization
is advantageous for sufficiently large $P$.\footnote{We note that all P\olya-Gamma sampling is done on CPU.} We also note that the computational complexity of wTGS is no worse than linear
in $N$, with the consequence that wTGS can be applied to datasets with large $N$ \emph{and} large $P$ in practice, at least if the sparsity assumption holds 
(i.e.~most variables are excluded in the posterior: $\gabs \ll P$).

\subsection{Correlated covariates scenario}
\label{sec:corr}

\begin{figure}[ht]
\centering
\begin{minipage}[b]{0.405\linewidth}
    \includegraphics[width=\linewidth]{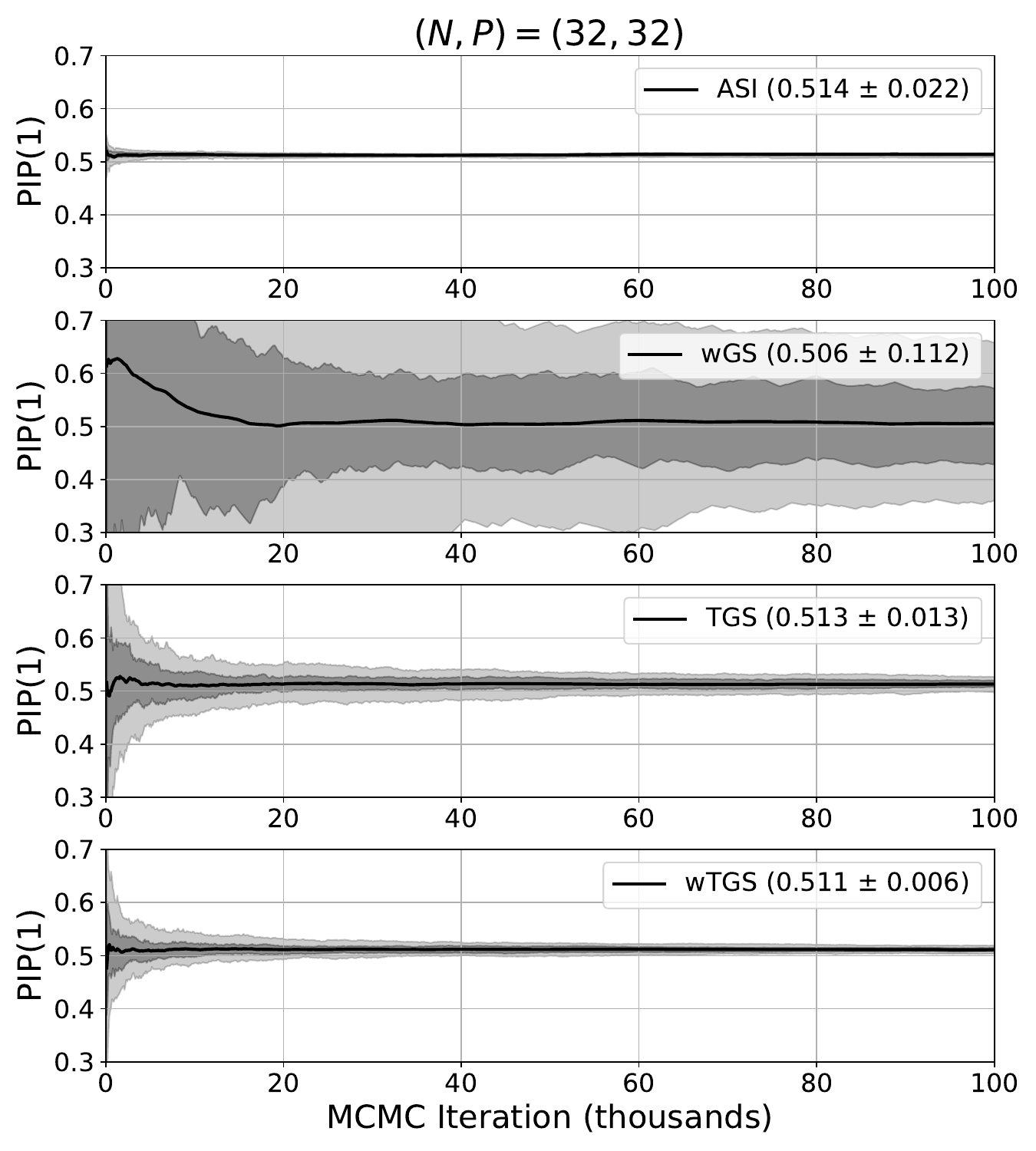}
\label{fig:corr32}
\end{minipage}
\quad
\begin{minipage}[b]{0.405\linewidth}
    \includegraphics[width=\linewidth]{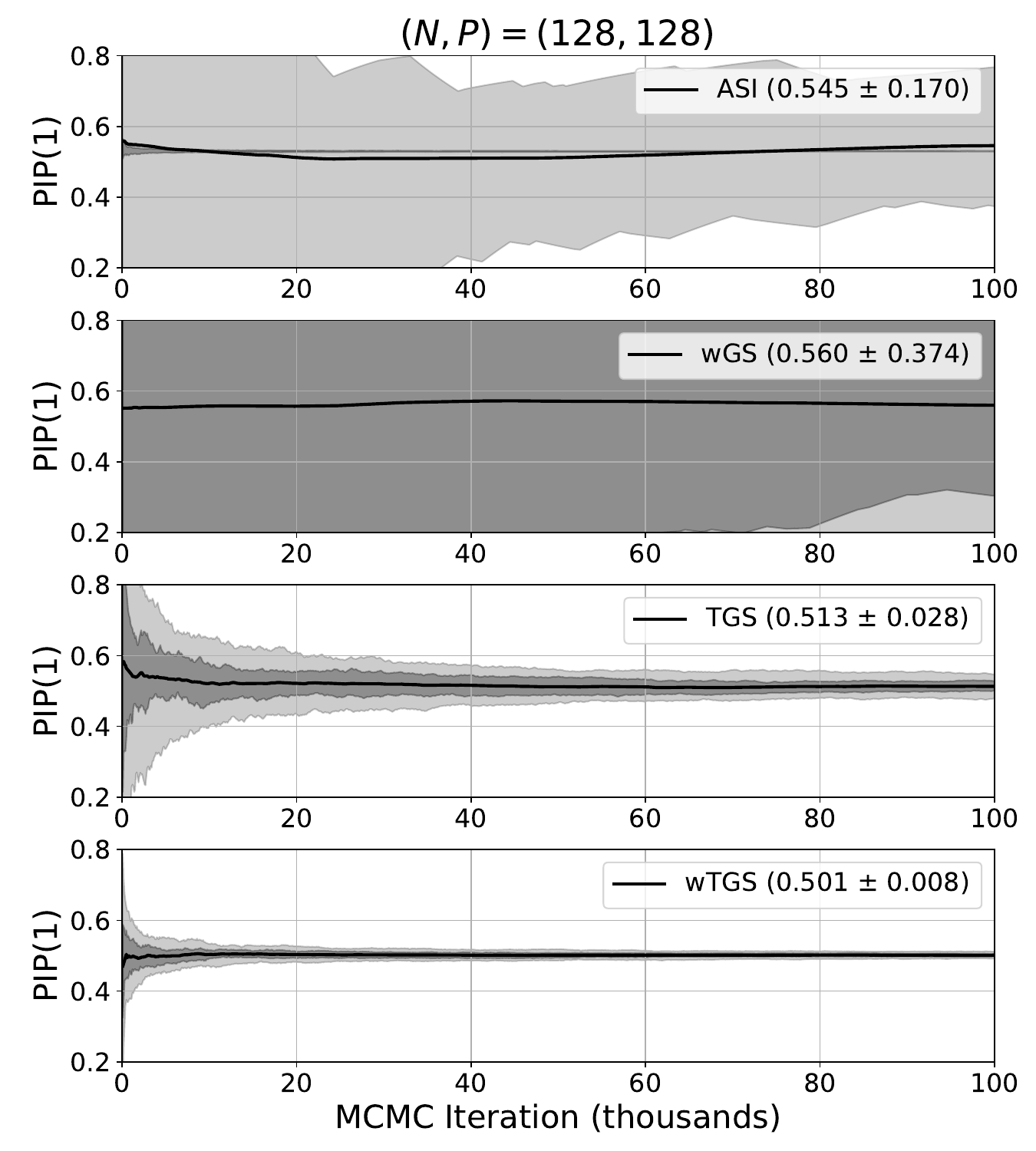}
\label{fig:corr128}
\end{minipage}
    \caption{We depict posterior inclusion probability (PIP) estimates for the first covariate in the scenario
    described in Sec.~\ref{sec:corr} for four different MCMC methods. 
    We depict results for $(N, P)=(32,32)$ [left] and $(N, P)=(128,128)$ [right].
    At each iteration $t$ the PIP is computed using all samples obtained through iteration $t$.
    The mean PIP is depicted with a solid black line and light and dark grey confidence intervals
    denote $10\%\!-\!90\%$ and $30\%\!-\!70\%$ quantiles, respectively.
    The true PIP is almost exactly $\tfrac{1}{2}$.
    In each case we run 100 independent chains. 
    For each method we also report the final PIP estimate (mean and standard deviation) in parentheses. 
    }
\label{fig:corr}
\end{figure}

\begin{figure}[ht]
    \includegraphics[width=0.245\linewidth]{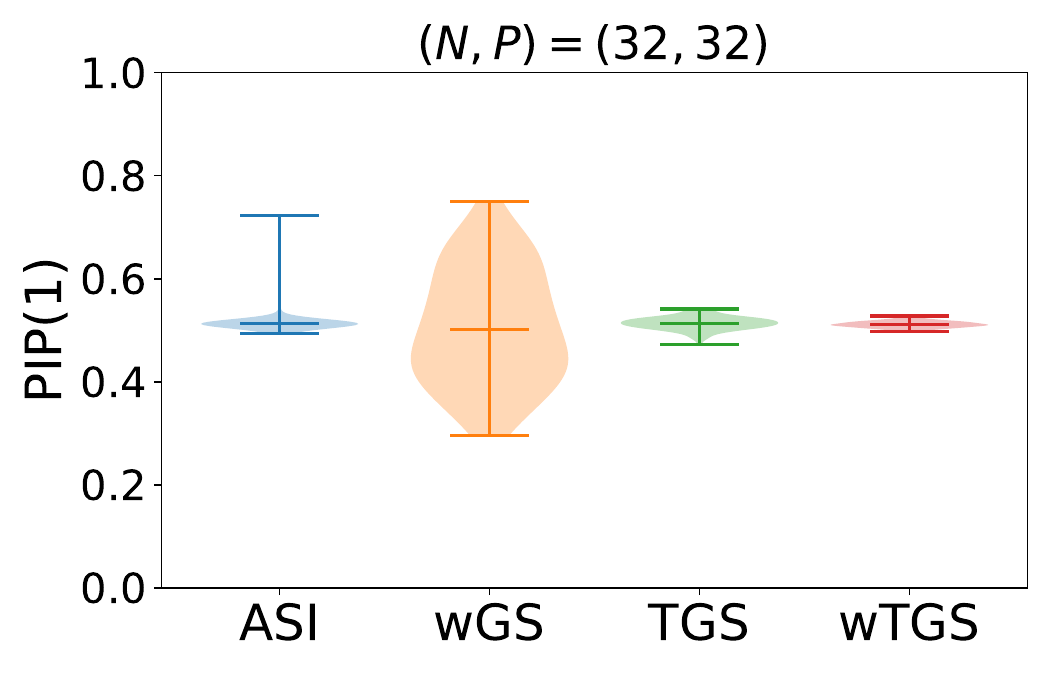}
    \includegraphics[width=0.245\linewidth]{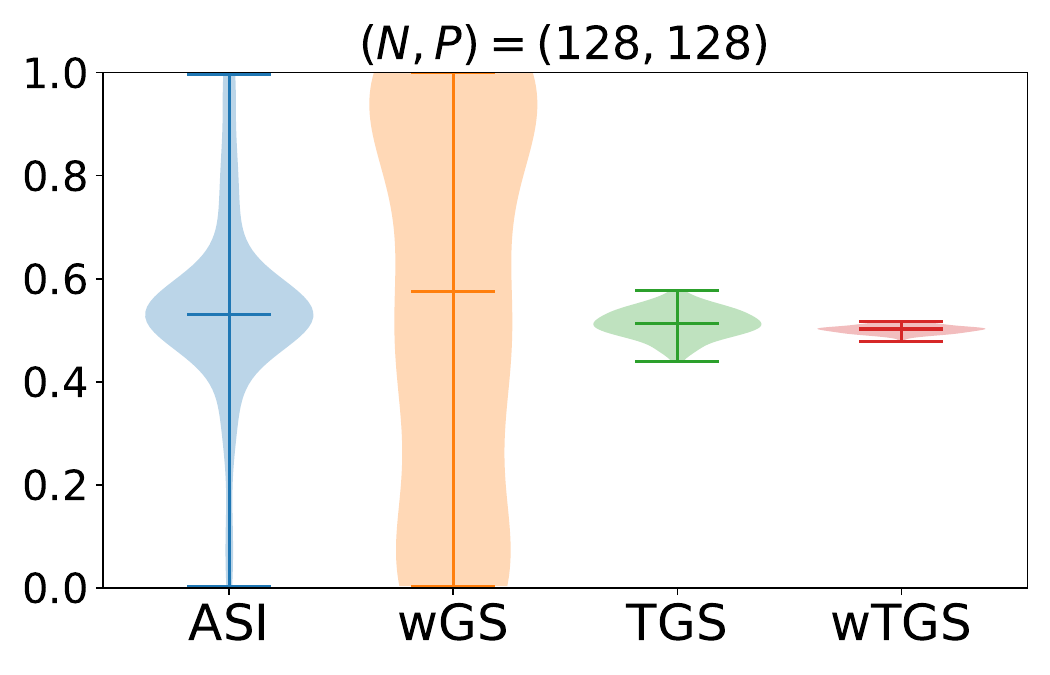}
    \includegraphics[width=0.245\linewidth]{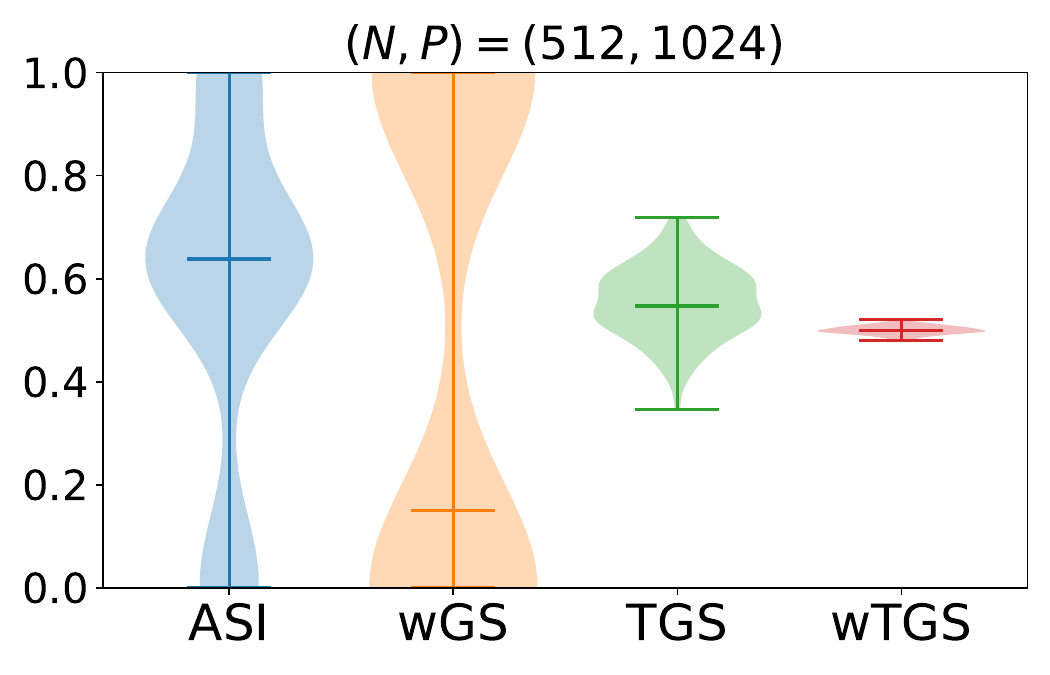}
    \includegraphics[width=0.245\linewidth]{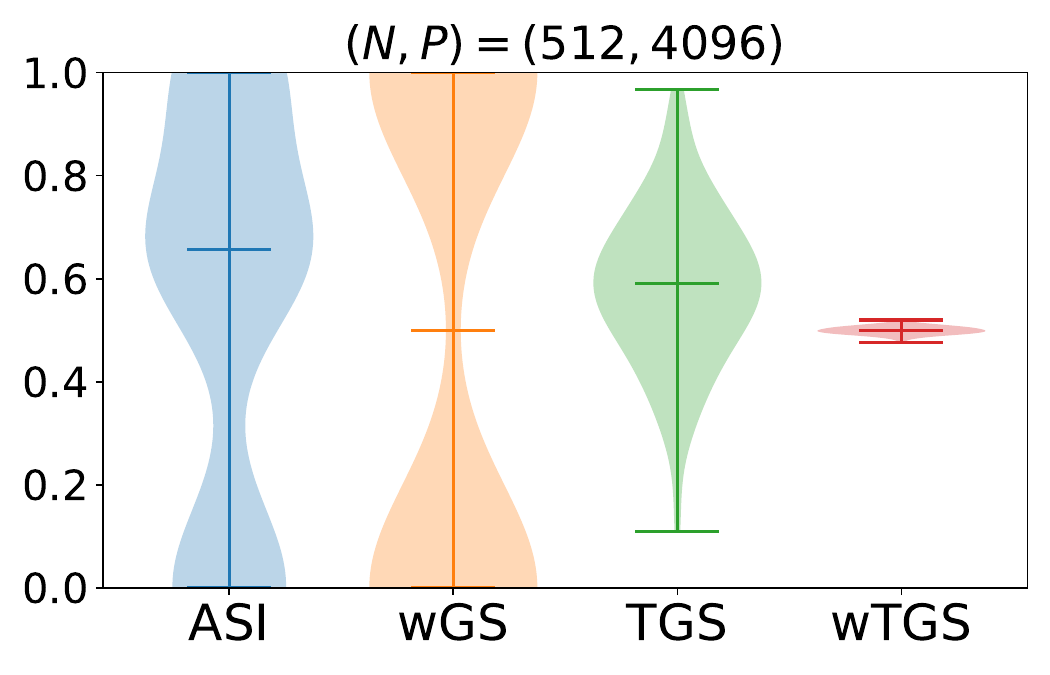}
    \caption{We depict violin plots for PIP estimates for the first covariate obtained with $10^5$ MCMC samples
    for four different methods on four datasets with varying numbers of data points $N$
    and covariates $P$. Horizontal bars denote the minimum, median, and maximum PIP estimates
    obtained from 100 independent MCMC runs. See Sec.~\ref{sec:corr} for details. 
    }
\label{fig:corrviolin}
\end{figure}

We consider simulated datasets in which two covariates ($i=1$ and $i=2$) are highly-correlated
and each on its own is able to fully explain the response $Y$. As discussed in
Sec.~\ref{sec:mg} this can be a challenging regime for MCMC methods, since it is easy for 
the MCMC chain to get stuck in one mode and fail to explore the other mode.
We consider four datasets with $(N, P) \in \{ (32, 32), (128, 128), (512, 1024), (512, 4096) \}$.
Each dataset has $\TC_n=10$ for all data points.
See Fig.~\ref{fig:corr} and Fig.~\ref{fig:corrviolin} for the results. 

We see that wGS does poorly on all datasets, including the smallest one with $P=32$ covariates.
By contrast wTGS yields low-variance PIP estimates in all cases. TGS does well for $P=32$ and
$P=128$ but exhibits large variance for $P\ge 1024$. Together these results demonstrate
the benefit of the non-sticky dynamics enabled by tempering, 
which allows (w)TGS to make frequent moves between
the two modes. Comparing wTGS to TGS we see that the targeted $i$-updates enabled by $\eta(\cdot)$ further
lower variance, especially for larger $P$. ASI estimates exhibit low variance
for $P=32$ (apart from a single outlier) but are high variance for larger $P$. This outcome is easy
to understand. Since ASI adapts its proposal distribution during warmup using a running estimate of 
each PIP, it is vulnerable to a rich-get-richer phenomenon in which covariates with
large initial PIP estimates tend to crowd out covariates with which they are highly correlated.
In the present scenario the result is that the ASI PIP estimates for the first two covariates 
are strongly anti-correlated. That this anti-correlation is ultimately due to suboptimal adaptation
is easily verified. For example for $P=1024$ ($P=4096$) the Pearson correlation coefficient
between the difference of final PIP estimates, i.e.~PIP(1) - PIP(2), and the difference of the
corresponding initial PIP estimates that define the proposal distribution is $0.904$ ($0.998$), respectively.

\subsection{Cancer data}
\label{sec:cancer}

We consider data from the Cancer Dependency Map project \citep{meyers2017computational,behan2019prioritization,pacini2021integrated,depmap1,depmap2}.
This dataset includes 900+ cancer cell lines spanning 26 different tissue types.
Each cell line has been subjected to a loss-of-function genetic screen that leverages CRISPR-Cas9
genetic editing to identify genes essential for cancer proliferation and survival. 
Genes identified by such screens are thought to be promising candidates for genetic vulnerabilities
that can be used to guide the development of treatment strategies
and novel therapeutics.

In more detail, we consider a subset of the data that includes $N=907$ cell lines
and $P=17273$ covariates, with each covariate encoding the RNA expression level for a given gene.
We consider two gene knockouts: DUSP4 and HNF1B. We make this choice because it is known
 that for both knockouts the RNA expression level of the corresponding gene is 
 highly predictive of cell viability---this serves as a useful sanity check of our results. 
 For each knockout the
dataset contains a real-valued response that encodes the effect size of knocking out 
that particular gene. This response variable has a heavy tail that skews left; this tail corresponds to the 
small number of cancer cell lines for which the gene knockout has a large detrimental effect on cell viability and reflects the significant variability among different cell lines.
We binarize this response variable by using the 20\% quantile as a cutoff.

\begin{figure}[ht]
    \includegraphics[width=0.49\linewidth]{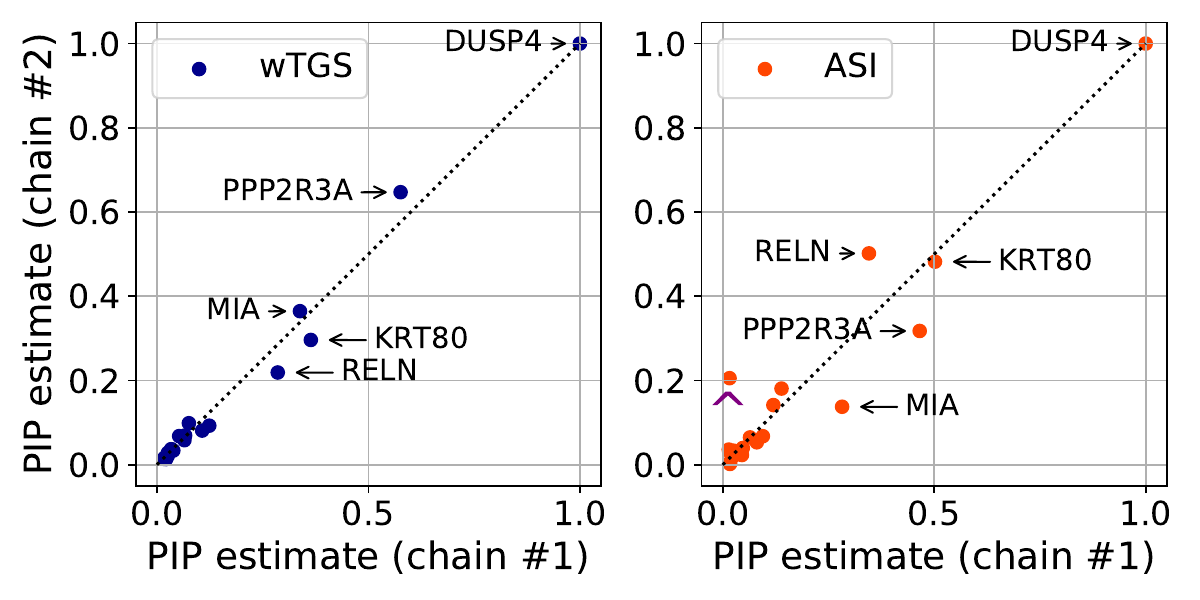}
    \includegraphics[width=0.49\linewidth]{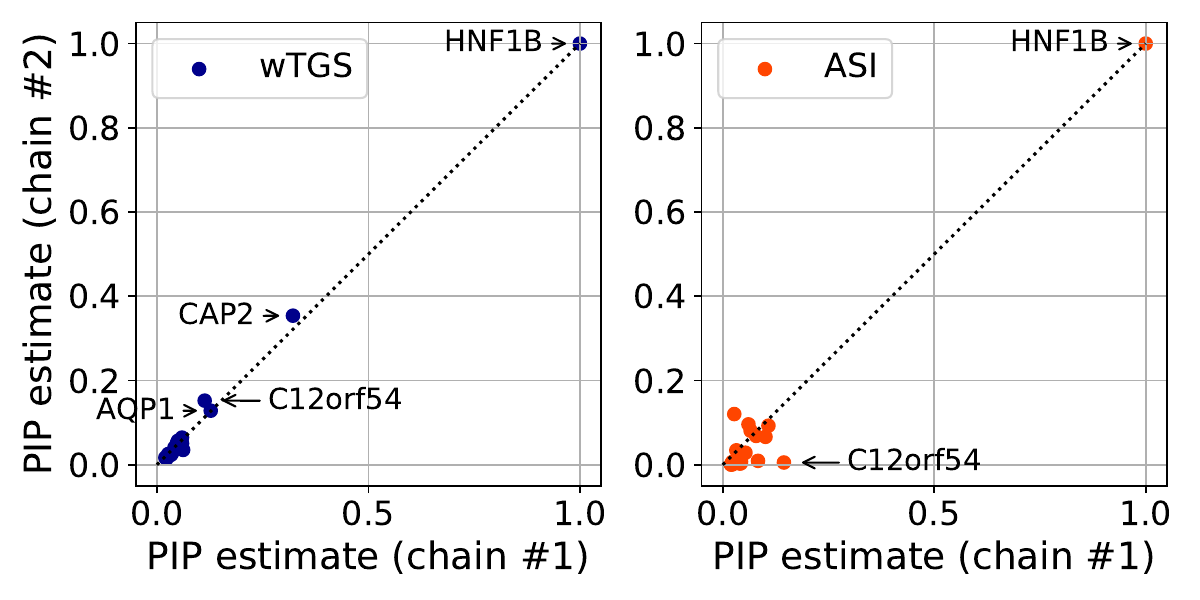}
    \caption{We depict PIP estimates for two independent MCMC chains 
    for two cancer datasets (left: DUSP4; right: HNF1B) using two MCMC methods. For each method we depict
    the top 20 PIPs from chain \#1 paired with the estimate from chain \#2. 
    The wTGS estimates show much better inter-chain concordance.
    For example, the PIPs obtained with ASI for KRT7 on the DUSP4 dataset
    (marked with a purple caret \^{}) differ by a factor of $12.8$ between the two chains,
    while the two wTGS estimates are $0.025$ and $0.021$.
    Similarly the PIPs obtained with ASI for C12orf54 on the HNF1B dataset
    differ by a factor of 26.9, while the two wTGS estimates are $0.113$ and $0.152$.
    See Sec.~\ref{sec:cancer} for details. 
    }
\label{fig:cancer}
\end{figure}

A minimum requirement for Bayesian variable selection to be useful in 
this setting is that results are reproducible across different MCMC runs.
In Fig.~\ref{fig:cancer} we depict the results we obtain from running pairs of independent
MCMC chains for $2.5 \times 10^5$ iterations after a burn-in of $2.5 \times 10^4$ iterations.
We see that the wTGS estimates show significantly more concordance between runs
than is the case for ASI. For example, if we look at genes for which the PIP estimate
for the DUSP4 dataset is at least $0.01$ ($0.10$) in at least one chain
and compute inter-chain PIP ratios,
then the maximum discrepancy for wTGS is $3.15$ ($1.33$), while 
 the maximum discrepancy for ASI is $76.77$ ($12.79$), respectively.
The corresponding numbers for HNF1B are $2.59$ ($1.35$) for wTGS and
$40.19$ ($26.91$) for ASI.

What makes this dataset particularly challenging for analysis is that the $P=17273$ covariates 
are strongly correlated. 
For example, the DUSP4 RNA expression level exhibits a correlation
greater than $0.40$ ($0.30)$ with $19$ ($203$) other covariates, respectively.
Indeed the MIA covariate identified by both algorithms (see Fig.~\ref{fig:cancer}) 
exhibits a $0.43$ correlation with DUSP4, while PPP2R3A
exhibits a $0.43$ correlation with KRT80.
Similarly the HNF1B RNA expression level has a correlation
greater than $0.70$ ($0.50)$ with $2$ ($33$) other covariates, respectively.
This is typical for these kinds of datasets, since similar cell types employ similar cellular `programs.'

\begin{figure}[ht]
    \includegraphics[width=0.49\linewidth]{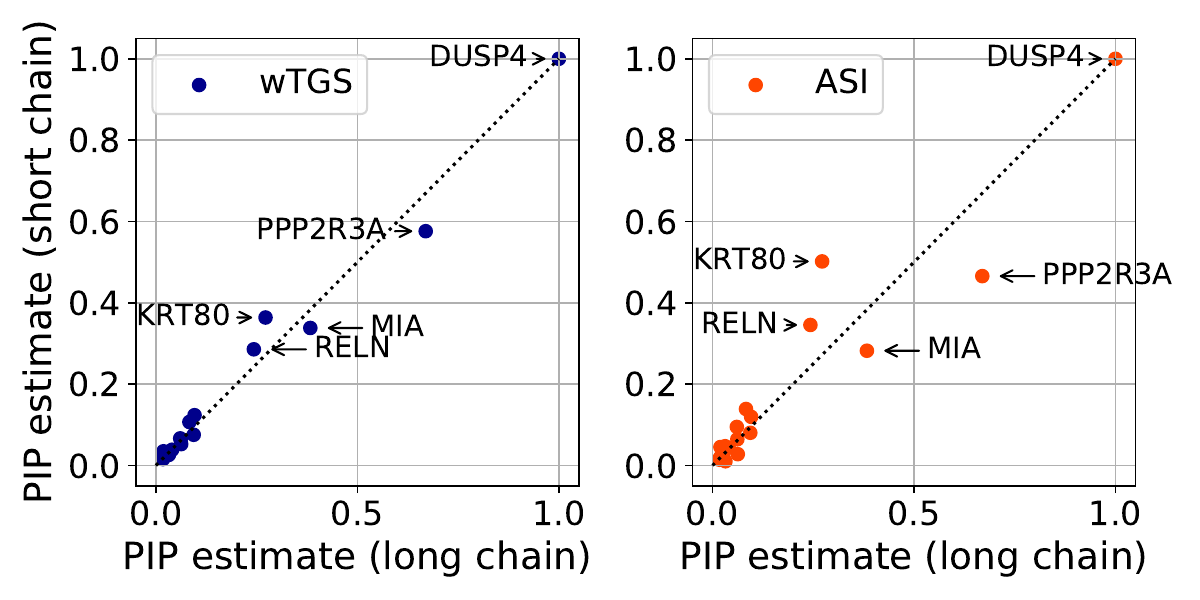}
    \includegraphics[width=0.49\linewidth]{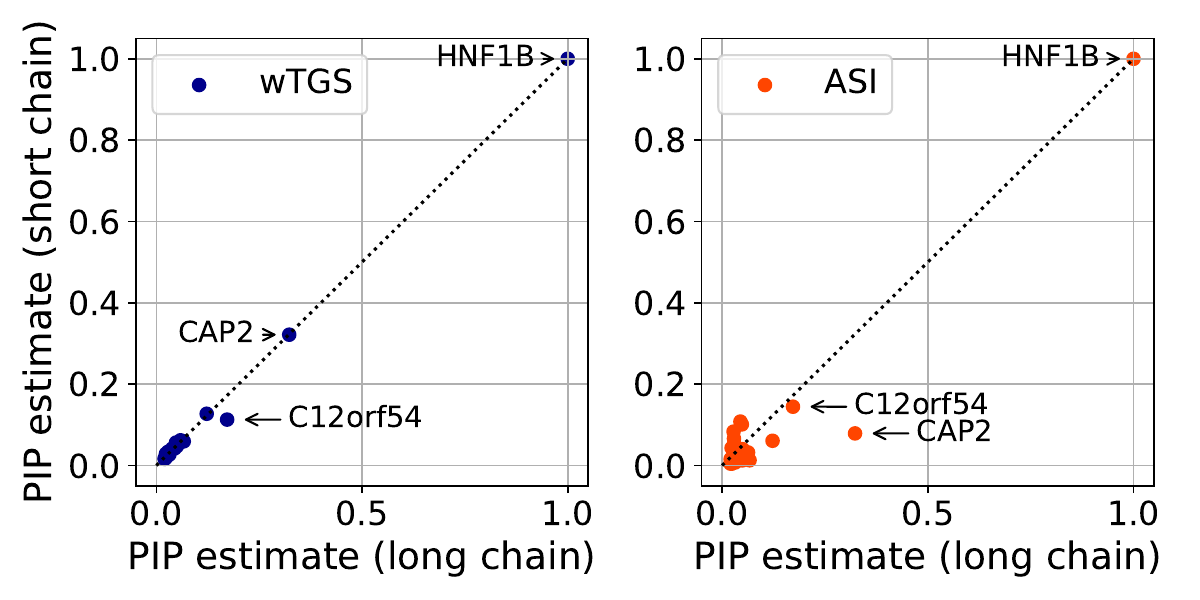}
    \includegraphics[width=0.49\linewidth]{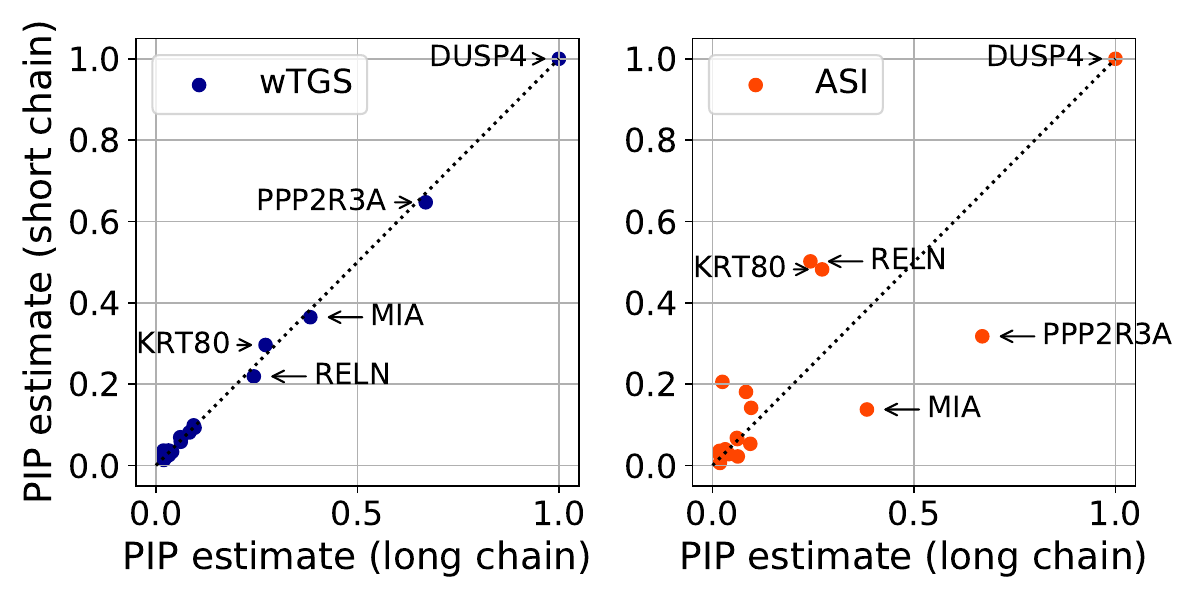}
    \includegraphics[width=0.49\linewidth]{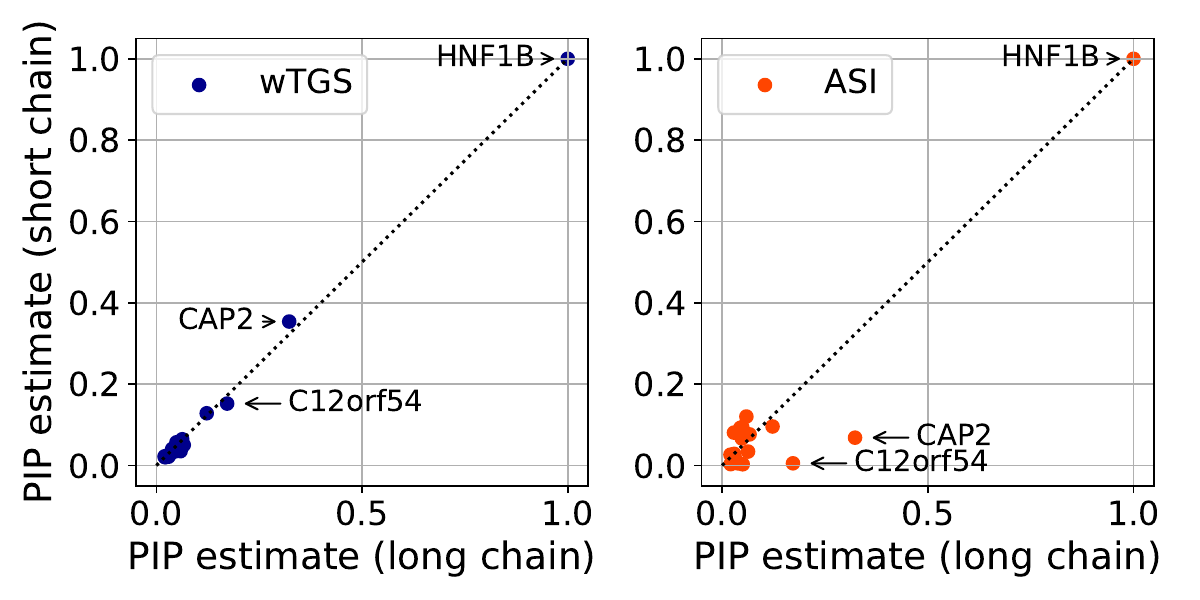}
    \caption{In this companion figure to Fig.~\ref{fig:cancer} we compare PIP estimates 
    obtained from short wTGS and ASI chains with $2.5 \times 10^5$ samples
    to a long wTGS chain with $5 \times 10^6$ samples.
    For each method we depict the top 20 PIPs from the long chain paired with estimates from
    the short chains (top row: chain \#1; bottom row: chain \#2).
    The wTGS estimates obtained with the short chains are significantly more accurate than is the case for ASI. 
    See Sec.~\ref{sec:cancer} for details. 
    }
\label{fig:cancerbig}
\end{figure}

The runtime for each wTGS chain in Fig.~\ref{fig:cancer} is $\sim\!70$ minutes on a NVIDIA Tesla K80 GPU.
To obtain a much lower variance result we run wTGS for $\sim\!24$ hours and obtain $5 \times 10^6$ samples.\footnote{Note that GPU utilization is low when running a single chain.
One could alternatively run 4 chains in parallel and obtain the same number of samples in about 8 hours.}
In Fig.~\ref{fig:cancerbig} we confirm that the good concordance exhibited
by wTGS PIP estimates obtained from chains with $2.5 \times 10^5$ samples reflects the
lower variance of wTGS PIP estimates as compared to ASI estimates.
See Table~\ref{table:cancer} in the supplemental materials for a list of all the top hits.

\subsection{Hospital data}
\label{sec:hospital}


We consider a hospital visit dataset with $N=1798$ considered in \citep{hilbe2011negative} and gathered
from Arizona Medicare data. The response variable is length of hospital stay for patients
undergoing a particular class of heart procedure and ranges between 1 and 53 days.
We expect the hospital stay to exhibit significant dispersion and so we use a negative binomial likelihood. 
There are three binary covariates: sex (female/male),
admission type (elective/urgent), and age (over/under 75). 
To make the analysis more challenging we add 97 superfluous covariates 
drawn i.i.d.~from a unit Normal distribution so that $P=100$.

\begin{figure}[ht]
    \centering
    \includegraphics[width=0.4\linewidth]{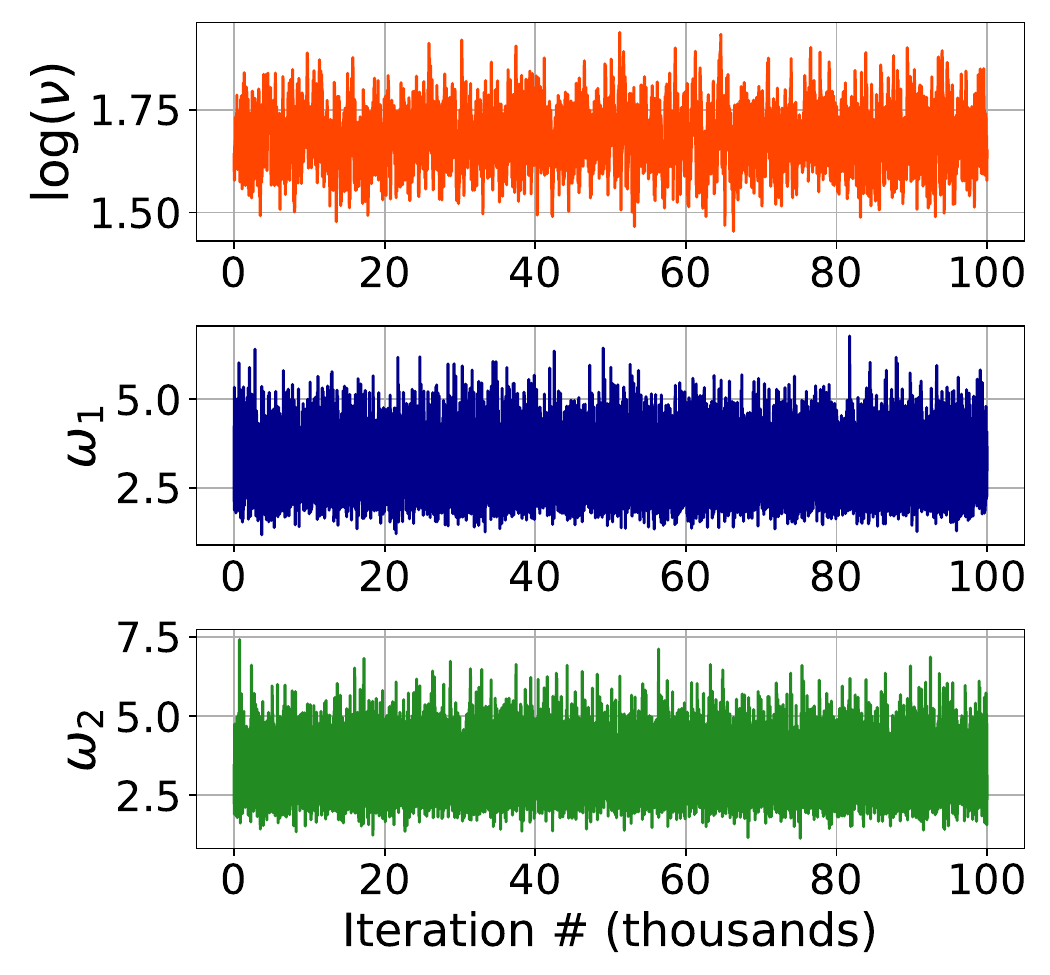}
    \includegraphics[width=0.4\linewidth]{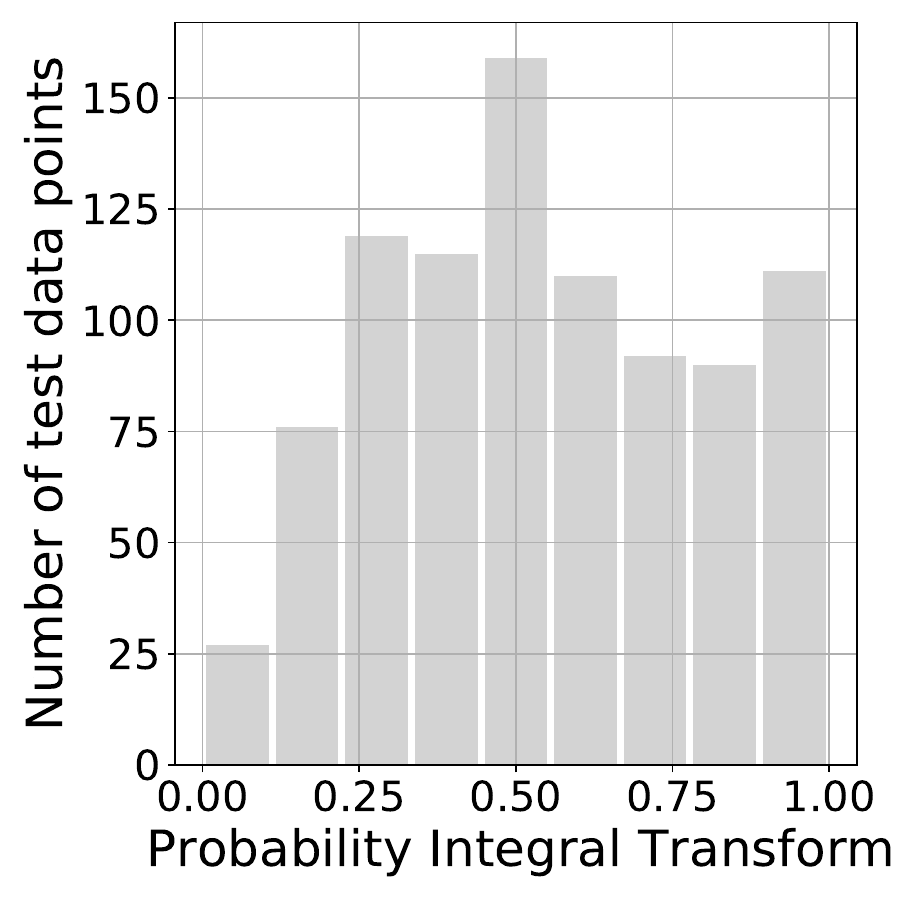}
    \caption{{\bf Left:} We depict trace plots for $\log(\nu)$ and two randomly chosen
    P\olya-Gamma variates $\omega_1$ and $\omega_2$ for a wTGS run on the hospital data 
    in Sec.~\ref{sec:hospital}. As expected the data is quite dispersed, with the posterior mean
    of the dispersion parameter $\nu$ being about $5.4$. {\bf Right:} We depict the Probability Integral Transform histogram
    for $899$ held-out test points for the hospital data in Sec.~\ref{sec:hospital}.
    }
\label{fig:nutraceplot}
\end{figure}

Running wTGS on the full dataset we find strong evidence for inclusion of two of the covariates: 
sex (PIP $\approx 0.95$) and admission type (PIP $\approx 1.0$). 
The corresponding coefficients are negative ($-0.15 \pm 0.02$) and positive ($0.63 \pm 0.03$), respectively.\footnote{
    Each estimate is conditioned upon inclusion of the corresponding covariate in the model.}
This corresponds to shorter hospital stays for males and longer hospital stays for urgent admissions.
In Fig.~\ref{fig:nutraceplot} (left) we depict trace plots for a few latent variables, each of which
is consistent with good mixing. 

Next we hold-out half of the dataset in order to assess the quality of the model-averaged predictive distribution.
We use the mean predicted hospital stay to rank the held-out patients and then partition them into two groups of equal
size. Comparing this predicted partition to the observed partition of patients into short- and long-stay patients,
we find a classification accuracy of 66.6\%. 
In Fig.~\ref{fig:nutraceplot} (right) we depict a more fine-grained predictive diagnostic, namely Dawid's 
Probability Integral Transform (PIT) \citep{dawid1984present}. Since the PIT values are approximately uniformly
distributed, we conclude that the predictive distribution is reasonably well-calibrated, although
probably somewhat overdispersed.

\subsection{Health survey data}
\label{sec:health}

We consider the German health survey with $N=1127$ considered in \citep{hilbe2007count}.
The response variable is the annual number of visits to the doctor
and ranges from $0$ to $40$ with a mean of $2.35$.
As in Sec.~\ref{sec:hospital}, we expect significant dispersion and thus use a negative binomial likelihood. 
There are two covariates: i) a binary covariate for self-reported health status (not bad/bad); and
ii) an age covariate, which ranges from 20 to 60.\footnote{We normalize the age covariate
so that it has mean zero and standard deviation one.}
To make the analysis more challenging we add 198 superfluous covariates 
drawn i.i.d.~from a unit Normal distribution so that $P=200$.

Running wTGS on the full dataset we find strong evidence for inclusion of the health status covariate
(PIP $\approx 1.0$) and only marginal evidence for including age (PIP $\approx 0.08$).
The health status coefficient is positive ($1.15 \pm 0.10$), suggesting
that patients whose health is self-reported as bad have $e^{1.15} \approx 3.17$ times as many visits to the doctor
as compared to those who report otherwise.\footnote{This is consistent with the raw empirical ratio, which is about $3.16$.} 
We find that the data are very overdispersed and infer the dispersion parameter to be $\nu = 0.99 \pm 0.07$.
See Sec.~\ref{app:nb} for additional details on wTGS for negative binomial regression.

\section{Discussion}
\label{sec:disc}

Several possible improvements to wTGS as outlined above suggest themselves.
A natural direction would be to combine the strengths of
wTGS and ASI.
Indeed one of the attractive features of ASI \citep{griffin2021search, wan2021adaptive}
is that the proposal enables moderately large moves in $\gamma$ space, i.e.~not just
single-site moves. Thus one possibility would be to incorporate such moves into wTGS,
perhaps interleaving them with the wTGS moves. Alternatively, one could make ASI
more like wTGS. In particular one could initialize the ASI proposal distribution
using PIP estimates obtained with wTGS. This should help alleviate the 
difficulty with highly-correlated covariates exhibited by ASI in experiments
in Sec.~\ref{sec:corr}-\ref{sec:cancer}.

An additional challenge is the binomial regression regime with large $N$ and large $\TC$. As mentioned in
Sec.~\ref{sec:tempinf}, $\omega$-updates may suffer from low acceptance probabilities
in this regime. Finding a suitably robust proposal distribution for this case
remains an open problem. 

More broadly, one of the principle challenges with Bayesian variable selection in
the highly-correlated setting beyond inference itself is that of interpretation.
How can one best convey to a user a manageable set of parsimonious 
models that are supported by the data? 
In scenarios with a large number of highly-correlated covariates, 
as is common for example in much of biology, posteriors that exhibit
marginal evidence for a large number of highly-correlated covariates 
are inherently difficult to interpret. While this issue is common to Bayesian
variable selection in all generalized
linear models and not just (negative) binomial regression, better methods
for making sense of complex high-dimensional posteriors are essential
if Bayesian model selection is to find wider use among data analysts and scientists alike.

\begin{ack}
We thank Jim Griffin for clarifying some of the details of the methodology described in reference \citep{wan2021adaptive} and Zolisa Bleki for help with the \texttt{polyagamma} package. 
We also thank James McFarland, Joshua Dempster, and Ashir Borah for help with DepMap data. 
\end{ack}

\bibliographystyle{plainnat}
\bibliography{ref}

\clearpage

\appendix
\section{Appendix}

This appendix is organized as follows.
In Sec.~\ref{app:mll} we discuss conditional marginal log likelihood computations.
In Sec.~\ref{app:cc} we discuss computational complexity.
In Sec.~\ref{app:omega} we discuss $\omega$-updates.
In Sec.~\ref{app:xi} we discuss $\xi$ adapation. 
In Sec.~\ref{app:nb} we discuss the modifications of our basic algorithm that are needed to accommodate negative binomial likelihoods.
In Sec.~\ref{app:fig} we include additional figures and tables accompanying Sec.~\ref{sec:exp}.
In Sec.~\ref{app:exp} we discuss experimental details.

\subsection{Efficient linear algebra for the (conditional) marginal log likelihood}
\label{app:mll}

\noindent The conditional marginal log likelihood $\log p(Y | X, \TC, \gamma, \omega)$ can be computed in closed form where, up to irrelevant constants, we have
\begin{align}
    \label{eqn:mll}
    \log p(Y | X, \TC, \gamma, \omega) &= \tfrac{1}{2} \ZZ_\gone \TT (X_\gone \TT \Omega X_\gone  + \tau \id_\gone)\inv \ZZ_\gone \\
    &-\tfrac{1}{2} \log \det(X_\gone \TT \Omega X_\gone   + \tau \id_\gone)    -\tfrac{1}{2} \log \det(\tau \inv \id_\gone)  \nonumber
\end{align}
where $\ZZ \in \RR^{P+1}$ with $\ZZ_j = \sum_{n=1}^N \kappa_n X_{n,j}$ 
for $j=1,...,P$ and the final component $\ZZ_{P+1} = \sum_{n=1}^N \kappa_n $ corresponds to the bias. Here and elsewhere in the appendix $X$ is augmented with a column of all ones
and $\kappa_n \equiv Y_n - \tfrac{1}{2} \TC_n$, 
$\Omega$ is the $N \times N$ diagonal matrix formed from $\omega$,  and $\gone$ is used to refer to the active
indices in $\gamma$ as well as the bias, which is always included in the model by assumption.
Using a Cholesky decomposition the quantity in Eqn.~\ref{eqn:mll} can be computed in $\OO(\gabs^3 + \gabs^2 N)$ time. 
If done naively this becomes expensive in cases where Eqn.~\ref{eqn:mll} needs to be computed for many values of $\gamma$ (as is needed e.g.~to compute Rao-Blackwellized PIPs). 
Luckily, and as is done by \citep{zanella2019scalable} and others in the literature, the computational cost can be
reduced significantly since we can exploit the fact that in practice we always consider `neighboring' values of $\gamma$
and so we can leverage rank-1 update structure where appropriate.
In the following we provide the formulae necessary for doing so. We keep the derivation generic
and consider the case of adding arbitrarily many variables to $\gamma$ even though in practice we only make use of the rank-1 formulae.

In more detail we proceed as follows.
Let $\II$ be the active indices in $\gamma$ together with the bias index $P+1$ (i.e.~we conveniently
augment $X$ by an all-ones feature column in the following). Let $\kk$ be a non-empty set of indices not in $\II$
and let $\II_\kk = \II \cup \kk$.
We let $\XX = \Omega^{\tfrac{1}{2}} X$
 and rewrite 
$F_\ik \equiv (\XX_\ik\TT \XX_\ik  + \tau \id_\ik)\inv$ in terms of $F_\II \equiv (\XX_\II \TT \XX_\II + \tau \id_\II)\inv$ as follows:
\begin{equation}
F_\ik =
  \begin{pmatrix}
  \XX_\II \TT \XX_\II  + \tau \id_\II& \XX_\II \TT \XX_\kk \\
  \XX_\kk \TT \XX_\II & \XX_\kk \TT \XX_\kk + \tau \id_\kk
  \end{pmatrix}\inv = 
    \begin{pmatrix}
  F_\II + F_\II \XX_\II \TT \XX_\kk  G_\kk \XX_\kk \TT X_\II   F_\II        &        -F_\II \XX_\II \TT \XX_\kk    G_\kk   \\
  -G_\kk   \XX_\kk \TT \XX_\II    F_\II                                                 &               G_\kk
  \end{pmatrix}
\end{equation}
where $G_\kk \inv \equiv \XX_\kk \TT \XX_\kk + \tau \id_\kk - \XX_\kk \TT \XX_\II F_\II \XX_\II \TT \XX_\kk$.

To efficiently compute the quadratic term in Eqn.~\ref{eqn:mll} we need
 to compute $\ZZ_\ik \TT F_\ik \ZZ_\ik $ in terms of $\ZZ_\II \TT F_\II \ZZ_\II$.
Write $\ZZ_\ik = (\ZZ_\II, \ZZ_\kk)$ so we have 
\begin{align}
\ZZ_\ik \TT F_\ik \ZZ_\ik =&
   \begin{pmatrix} \ZZ_\II \\ \ZZ_\kk \end{pmatrix} \TT
    \begin{pmatrix}
  F_\II + F_\II \XX_\II \TT \XX_\kk  G_\kk \XX_\kk \TT \XX_\II   F_\II        &        -F_\II \XX_\II \TT \XX_\kk    G_\kk   \\
  -G_\kk   \XX_\kk \TT \XX_\II    F_\II                                                 &               G_\kk
  \end{pmatrix}
  \begin{pmatrix} \ZZ_\II \\ \ZZ_\kk \end{pmatrix} \\
  =\;& \ZZ_\II \TT F_\II \ZZ_\II + \ZZ_\II \TT F_\II \XX_\II \TT \XX_\kk  G_\kk \XX_\kk \TT \XX_\II   F_\II  \ZZ_\II  \\
   &- \ZZ_\II \TT F_\II \XX_\II \TT \XX_\kk    G_\kk   \ZZ_\kk - \ZZ_\kk \TT G_\kk   \XX_\kk \TT \XX_\II    F_\II  \XX_\II    +  \ZZ_\kk \TT G_\kk   \ZZ_\kk   \\
   =\;& \ZZt_\II \TT \ZZt_\II  + (\XXt_\kk \TT \XXt_\II \ZZt_\II )\TT (\XXt_\kk \TT \XXt_\II \ZZt_\II ) -2 (\XXt_\kk \TT \XXt_\II \ZZt_\II )\TT \ZZt_\kk +  \ZZt_\kk \TT \ZZt_\kk  \\
   =\;& \ZZt_\II \TT \ZZt_\II  + \left \vert \left\vert \XXt_\kk \TT \XXt_\II \ZZt_\II - \ZZt_\kk  \right\vert\right\vert^2. 
\end{align}
where $ \left \vert \left\vert \cdot  \right\vert\right\vert$ is the 2-norm in $\RR^{|\kk|}$ and we define
\begin{align}
\label{eqn:mllchols}
L_\II  L_\II \TT&= \XX_\II \TT \XX_\II + \tau \id_\II= F_\II \inv      \\ 
L_\kk  L_\kk \TT &= \XX_\kk \TT \XX_\kk + \tau \id_\kk - \XX_\kk \TT \XX_\II F_\II \XX_\II \TT \XX_\kk 
                             = \XX_\kk \TT \XX_\kk + \tau \id_\kk  - \XX_\kk \TT \XXt_\II \XXt_\II \TT \XX_\kk  = G_\kk \inv \nonumber \\
\ZZt_\II &\equiv L_\II \inv \ZZ_\II \qquad \qquad        \ZZt_\kk \equiv L_\kk \inv \ZZ_\kk \qquad       \XXt_\II \equiv  \XX_\II L_\II \invT   \qquad     \XXt_\kk \equiv  \XX_\kk L_\kk \invT \nonumber
\end{align}
Here $L_\II$ and $L_\kk$ are Cholesky factors.
This can be rewritten as
\begin{align}
\ZZ_\ik \TT F_\ik \ZZ_\ik =  \ZZt_\II \TT \ZZt_\II  + \left \vert \left\vert  W_\kk  \right\vert\right\vert^2
\qquad {\rm with} \qquad   W_\kk \equiv L_\kk \inv  \left(  \XX_\kk \TT  \XXt_\II \ZZt_\II - \ZZ_\kk  \right)
\end{align}
Together these formulae can be used to compute the quadratic term efficiently.

Next we turn to the log determinant in Eqn.~\ref{eqn:mll}.
We begin by noting that 
\begin{align}
 \log \det \left( \XX_\ik \TT \XX_\ik + \tau \id_\ik \right) + \log\det( \tau \inv \id_\ik) =  \log\det(\Omega) + \log \det \left(X_\ik X_\ik\TT/\tau +\Omega\inv \right)
\end{align}
and
\begin{align}
\log \det \left(X_\ik X_\ik\TT/\tau +\Omega\inv \right) &= \log \det \left(X_\II X_\II \TT/\tau +\Omega\inv \right) + 
\log \det \left( \id_\kk/\tau\right) \\ &+ \log \det \left(\tau\id_\kk + X_\kk \TT (X_\II X_\II \TT/\tau +\Omega\inv)\inv X_\kk \right)
\end{align}
which together imply
\begin{align}
\{  \log \det \left( \XX_\ik \TT \XX_\ik + \tau \id_\ik \right) + \log\det( \tau \inv \id_\ik)  \} - \nonumber \\ \nonumber
\{  \log \det \left( \XX_\II \TT \XX_\II + \tau \id_\II \right) + \log\det( \tau \inv \id_\II)  \} &= \nonumber
\log \det \left(X_\ik X_\ik\TT/\tau +\Omega\inv \right) - \log \det \left(X_\II X_\II \TT/\tau +\Omega\inv \right) \\
&=  \nonumber
 \log \det \left(\id_\kk +  \tau\inv X_\kk \TT (X_\II X_\II \TT/\tau +\Omega\inv)\inv X_\kk \right)
\end{align}
While these equations can be used to compute the log determinant reasonably efficiently, they 
exhibit cubic computational complexity w.r.t.~$N$. So instead we write
\begin{align}
\det( \XX_\ik \TT \XX_\ik  + \tau \id_\ik) &= \det
  \begin{pmatrix}
  \XX_\II \TT \XX_\II  + \tau \id_\II& \XX_\II \TT \XX_\kk \\
  \XX_\kk \TT \XX_\II & \XX_\kk \TT \XX_\kk + \tau \id_\kk
  \end{pmatrix} \\ 
  &= \det(\XX_\kk \TT \XX_\kk + \tau \id_\kk -  \XX_\kk \TT \XX_\II  ( \XX_\II \TT \XX_\II  + \tau \id_\II)\inv \XX_\II \TT \XX_\kk ) \times \det( \XX_\II \TT \XX_\II  + \tau \id_\II) \nonumber \\
  &= \det( G_\kk \inv ) \times \det( \XX_\II \TT \XX_\II  + \tau \id_\II)   \nonumber
\end{align}
This form is convenient because it relies on the term $G_\kk$ that we in any case need to compute the quadratic form. Similarly $ \det( \XX_\II \TT \XX_\II  + \tau \id_\II)$ is easily computed
from the Cholesky factor $L_\II$.

\subsection{Computational complexity}
\label{app:cc}

The primary computational cost in all four algorithms considered in the main text
arises in computing the conditional PIPs $p(\gamma_j = 1 | \gmj, \omega) $ for $j=1,...,P$.
The MCMC algorithms wTGS, TGS, and ASI all use $p(\gamma_j = 1 | \gmj, \omega)$ directly in
that (conditional) PIPs are used to define the Markov chain, although in the case of ASI the PIPs
are only strictly necessary during the warm-up/adaptation phase. In any case computing
these quantities is necessary for Rao-Blackwellizing PIP estimates, and so in
practice all four algorithms spend the majority of compute on calculating $p(\gamma_j = 1 | \gmj, \omega)$. 
The next largest computational cost is usually sampling P\olya-Gamma variables, although
this is $\OO(N)$ and so the cost is moderate in most cases. For wTGS, TGS, and wGS 
computing the acceptance probability Eqn.~\ref{eqn:alphaomega} is another subdominant but non-negligible cost.

The precise computational cost of computing $p(\gamma_j = 1 | \gmj, \omega)$ depends on the details
of how the formulae in Sec.~\ref{app:mll} are implemented. For example in \citep{zanella2019scalable},
where there is no $\omega$ variable, additional computational savings are made possible by caching
certain intermediate quantites from iteration to iteration. (Additionally, in this setting it can be
advantageous to pre-compute $X \TT X$.) In our case where $\omega$ changes
every few MCMC iterations, there is little to be gained from the additional complexity that would
be required by such caching. Instead we compute most quantities afresh from iteration to iteration.

Using the rank-1 update formulae from Sec.~\ref{app:mll} the result is $\OO(\gabs N P + N \gabs^3 + \gabs^4)$ computational complexity per MCMC iteration with $i>0$ and $\OO(N)$ per MCMC iteration with $i=0$.
Here the $\OO(\gabs N P)$ term arises from computing $\XXt_\II \TT \XX_\kk$ for $\gabs$ active indices and $\OO(P)$ inactive
indices $\kk$ (see Eqn.~\ref{eqn:mllchols}).
We note, however, that these asymptotic formulae are somewhat misleading in practice, since
most of the necessary tensor ops are highly-parallelizable and very efficiently implemented on modern hardware. 
For this reason Fig.~\ref{fig:runtime} is particularly useful for understanding the runtime in practice, since
the various parts of the computation will be more or less expensive depending on the precise regime and the underlying
low-level implementation and hardware.
Here the $\OO(N \gabs^3 + \gabs^4)$ terms come from our avoidance of rank-1 downdate formulae: in essence
we compute $p(\gamma_j = 1 | \gmj, \omega)$ directly without any tricks for the (assumed small) number of $j$
with $\gamma_j = 1$.
We note that the computational complexity could be improved at the cost of additional code complexity and possible
round-off errors.

\subsection{$\omega$-update}
\label{app:omega}

The acceptance probability for the $\omega$-update in Sec.~\ref{sec:tempinf} is given by
\begin{align}
\label{eqn:appalphaomega}
\alpha(\omega  \! \to \! \omega^\prime | \gamma) &=  \min\left(1,
      \frac{p(Y | \gamma, \omega^\prime, X, \TC) p(\gamma) p(\omega^\prime|\TC) }{p(Y | \gamma, \omega, X, \TC) p(\gamma) p(\omega|\TC)}
      \frac{p(\omega | \gamma, \bhat(\gamma, \omega^\prime), \DD)}{p(\omega^\prime | \gamma, \bhat(\gamma, \omega), \DD)} \right) 
\end{align}
where the ratio of proposal densities is given by
\begin{align}
 \frac{p(\omega | \gamma, \bhat(\gamma, \omega^\prime), \DD)}{p(\omega^\prime | \gamma, \bhat(\gamma, \omega), \DD)} &=
          \frac{p(Y | \gamma, \omega,  \bhat(\gamma, \omega^\prime), X, \TC) p(\gamma) p(\omega|\TC) p(\bhat(\gamma, \omega^\prime))}
      {\int \! d \omhat \, p(Y | \gamma, \omhat,  \bhat(\gamma, \omega^\prime), X, \TC) p(\gamma) p(\omhat|\TC) p(\bhat(\gamma, \omega^\prime))} \times \nonumber \\
           & \left\{ \frac{p(Y | \gamma, \omega^\prime,  \bhat(\gamma, \omega), X, \TC) p(\gamma) p(\omega^\prime|\TC) p(\bhat(\gamma, \omega))}
      {\int \! d \omhat \, p(Y | \gamma, \omhat,  \bhat(\gamma, \omega), X, \TC) p(\gamma) p(\omhat|\TC) p(\bhat(\gamma, \omega))} \right \} \inv  
\end{align}
Simplifying we have that the ratio in $\alpha(\omega  \! \to \! \omega^\prime | \gamma)$ is given by
\begin{align}
    &\frac{p(Y | \gamma, \omega^\prime, X, \TC) }{p(Y | \gamma, \omega, X, \TC)}
          \frac{p(Y | \gamma, \omega,  \bhat(\gamma, \omega^\prime), X, \TC)}
      {\int \! d \omhat \, p(Y | \gamma, \omhat,  \bhat(\gamma, \omega^\prime), X, \TC) p(\omhat|\TC)}
            \frac{\int \! d \omhat \, p(Y | \gamma, \omhat,  \bhat(\gamma, \omega), X, \TC) p(\omhat|\TC)}{p(Y | \gamma, \omega^\prime,  \bhat(\gamma, \omega), X, \TC) }
\nonumber \\
&= \frac{p(Y | \gamma, \omega^\prime, X, \TC) }{p(Y | \gamma, \omega, X, \TC)}
          \frac{p(Y | \gamma, \omega,  \bhat(\gamma, \omega^\prime), X, \TC)}
      {p(Y | \gamma, \bhat(\gamma, \omega^\prime), X, \TC) }
            \frac{ p(Y | \gamma, \bhat(\gamma, \omega), X, \TC)}{p(Y | \gamma, \omega^\prime,  \bhat(\gamma, \omega), X, \TC) }
\nonumber \\
&= \frac{p(Y | \gamma, \omega^\prime, X, \TC) }{p(Y | \gamma, \omega, X, \TC)}
\frac{p(Y | \gamma, \omega, \bhat(\gamma, \omega^\prime), X, \TC) }{p(Y | \gamma, \omega^\prime, \bhat(\gamma, \omega), X, \TC)}
\frac{p(Y | \gamma,  \bhat(\gamma, \omega), X, \TC) }{p(Y | \gamma, \bhat(\gamma, \omega^\prime), X, \TC)} \nonumber
\end{align}
which is Eqn.~\ref{eqn:alphaomega} in the main text.
Here
\begin{align}
\label{eqn:alphaomega2}
\frac{p(Y | \gamma, \omega, \bhat(\gamma, \omega^\prime), X, \TC) }{p(Y | \gamma, \omega^\prime, \bhat(\gamma, \omega), X, \TC)} &=
\frac{\exp(\kappa \cdot \psihat(\gamma, \omega^\prime) -\tfrac{1}{2} \omega \cdot \psihat(\gamma, \omega^\prime)^2) }{\exp(\kappa \cdot \psihat(\gamma, \omega) -\tfrac{1}{2} \omega^\prime \cdot \psihat(\gamma, \omega)^2)}
\end{align}
and
\begin{align}
\frac{p(Y | \gamma,  \bhat(\gamma, \omega), X, \TC) }{p(Y | \gamma, \bhat(\gamma, \omega^\prime), X, \TC)} &=
\frac{\prod_n \exp(\psihat(\gamma, \omega)_n))^{Y_n}}{\prod_n (1 + \exp(\psihat(\gamma, \omega)_n))^{\TC_n}}
\frac{\prod_n (1 + \exp(\psihat(\gamma, \omega^\prime)_n))^{\TC_n}}{\prod_n \exp(\psihat(\gamma, \omega^\prime)_n))^{Y_n}}
\end{align}
where
\begin{align}
(\psihat(\gamma, \omega))_n \equiv \bhat(\gamma, \omega)_0 + \bhat(\gamma, \omega)_\gamma  \cdot X_{n\gamma}
\end{align}
and
\begin{align}
    \bhat(\gamma, \omega) = (X_\gone \TT \Omega X_\gone + \tau \mathbb{1}_\gabsone)\inv X_\gone \TT \kappa
    \in \RR^{\gabsone}
    \label{eqn:bhatdefn}
\end{align}
where as in Sec.~\ref{app:mll} $X$ is here augmented with a column of all ones.
As detailed in \citep{polson2013bayesian} the (approximate) Gibbs proposal distribution
that results from conditioning on $\bhat$ is given by a P\olya-Gamma distribution determined by $\TC$ and $\psihat$:
\begin{align}
    p(\omega^\prime | \gamma, \bhat(\gamma, \omega), \DD) = {\rm PG}(\omega^\prime | \TC, \psihat(\gamma, \omega))
\end{align}
In practice we do without the MH rejection step for $\omega$ in the early stages of burn-in to allow the 
MCMC chain to more quickly reach probable states.

\subsection{$\xi$-adaptation}
\label{app:xi}

The magnitude of $\xi$ controls the frequency of $\omega$ updates. Ideally
$\xi$ is such that an $\OO(1)$ fraction of MCMC iterations result in a $\omega$ update,
with the remainder of the computational budget being spent on $\gamma$ updates.
Typically this can be achieved by choosing $\xi$ in the range $\xi \sim 1-5$.
Here we describe a simple scheme for choosing $\xi$ adaptively 
during burn-in to achieve the desired behavior.

We introduce a hyperparameter $f_\omega \in (0, 1)$ that controls the desired $\omega$
update frequency. Here $f_\omega$ is normalized such that $f_\omega=1$ corresponds to a situation
in which all updates are $\omega$ updates, i.e.~all states in the MCMC chain are in the $i=0$ state
(something that would be achieved by taking $\xi \to \infty$). Since $\omega$ updates are of somewhat
less importance for obtaining accurate PIP estimates than $\gamma$ updates, we recommend a somewhat
moderate value of $f_\omega$, e.g. $f_\omega \sim 0.1-0.4$. 
For all experiments in this paper we use $f_\omega = 0.25$.

Our adaptation scheme proceeds as follows.
We initialize $\xi_0 = 5$. At iteration $t$ during the burn-in a.k.a.~warm-up phase
we update $\xi_t$ as follows: 
\begin{align}
    \xi_{t+1} = \xi_t +  \frac{f_\omega - \frac{\xi_t}{\phi(\gamma_t, \omega_t)} }{\sqrt{t + 1}}
\end{align}
By construction this update aims to achieve that a fraction $f_\omega$ of MCMC states
satisfy $i=0$, since the quantity
\begin{align}
\phi(\gamma, \omega) = \xi +  \frac{1}{P} \sum_{i=1}^P  \frac{\tfrac{1}{2} \eta(\gmi, \omega)}{ p(\gamma_i | \gmi, \omega, \DD)}
\end{align}
encodes the total probability mass assigned to states $i=0$ and $i>0$.

\subsection{Negative binomial likelihood}
\label{app:nb}

We specify in more detail how we can accommodate the negative binomial likelihood using
P\olya-Gamma augmentation. Using the identity Eqn.~\ref{eqn:pgidentity} we write
\begin{align}
    {\rm NegBin}(Y_n | \psi_n, \nu) &= \frac{\Gamma(Y_n + \nu)}{\Gamma(Y_n + 1)\Gamma(\nu)}
    \left(\frac{\exp(\psi_n + \psi_0 - \log \nu)}{1 + \exp(\psi_n + \psi_0 - \log \nu)} \right)^{Y_n}
    \left(\frac{1}{1 + \exp(\psi_n + \psi_0- \log \nu)} \right)^{\nu} \\
    &\propto \frac{1}{2^{\nu}} e^{\tfrac{1}{2}(Y_n - \nu)(\psi_n + \psi_0- \log \nu)} 
    \EE_{p(\omega_n |Y_n + \nu, 0)} \left[ \exp(-\tfrac{1}{2} \omega_n (\psi_n + \psi_0 - \log \nu)^2) \right] \nonumber
\end{align}
where as before $\psi_n = \beta_0 + \beta_\gamma  \cdot X_{n\gamma}$
and $\psi_0$ is a user-specified offset. Here $\nu >0$ controls the overdispersion of the negative binomial likelihood.
We note that by construction the mean of ${\rm NegBin}(Y_n | \psi_n, \nu)$ is given by $\exp(\psi_n + \psi_0)$.
Thus $\psi_0$ (which can potentially depend on $n$) can be used to specify a prior mean for $Y$.
This is equivalent to adjusting the prior mean of the bias $\beta_0$.

Comparing to Sec.~\ref{sec:pg} we see that $\kappa_n$ is now given by $\kappa_n = \tfrac{1}{2}(Y_n - \nu)$.
When computing $\log p(Y | X, \gamma, \omega)$ the quantity $\ZZ$ now becomes 
$\ZZ_j = \sum_n X_{n,j} \left(\kappa_n - \omega_n(\psi_0 - \log \nu) \right)$, see Sec.~\ref{app:mll}. 
One also picks up an additional factor of 
\begin{align}
    \exp(\kappa \cdot (\psi_0 - \log \nu) - \tfrac{1}{2} \omega \cdot (\psi_0 - \log \nu)^2) 
\end{align}

In our experiments we infer $\nu$, which we assume to be unknown. For simplicity we put a flat (i.e.~improper)
prior on $\log \nu$, although other choices are easily accommodated. To do so we modify the $\omega$ update
described in Sec.~\ref{app:omega} to a joint $(\omega, \log \nu)$ update. In more detail we use a simple
gaussian random walk proposal for $\log \nu$ with a user-specified scale (we use $0.03$ in our experiments).
Conditioned on a proposal $\log \nu^\prime$ we then sample a proposal $\omega^\prime$. Similar
to the binomial likelihood case, we do this by computing
\begin{align}
    \bhat(\gamma, \omega, \nu) \equiv \EE_{p(\beta | \gamma, \omega, \nu, \DD)} \left[ \beta \right]
\end{align}
and use a proposal distribution $\omega^\prime \sim p(\cdot | \gamma, \bhat, \nu^\prime, \DD)$.
In the negative binomial case the formula for $\bhat$ in Eqn.~\ref{eqn:bhatdefn} becomes
\begin{align}
    \bhat(\gamma, \omega, \nu) = (X_\gone \TT \Omega X_\gone + \tau \mathbb{1}_\gabsone)\inv X_\gone \TT
    \left(\kappa - \omega(\psi_0 - \log \nu) \right)
    \in \RR^{\gabsone}
\end{align}
Additionally the proposal distribution is given by 
\begin{align}
    p(\omega^\prime | \gamma, \bhat(\gamma, \omega, \nu), \nu^\prime, \DD) = 
    {\rm PG}(\omega^\prime | Y + \nu^\prime, \psihat(\gamma, \omega, \nu))
\end{align}
The acceptance probability can then be computed as in Sec.~\ref{app:omega}, although in this 
case the resulting formulae are somewhat more complicated because of the need to 
keep track of $\nu$ and $\nu^\prime$ as well as the fact that there is less scope for cancellations
so that we need to compute quantities like $\Gamma(\nu)$. Happily, just like in the binomial regression case,
the acceptance probability can be computed without recourse to the P\olya-Gamma density.

\subsection{Additional figures and tables}
\label{app:fig}

\begin{figure}[ht]
\centering
\begin{minipage}[b]{0.405\linewidth}
    \includegraphics[width=\linewidth]{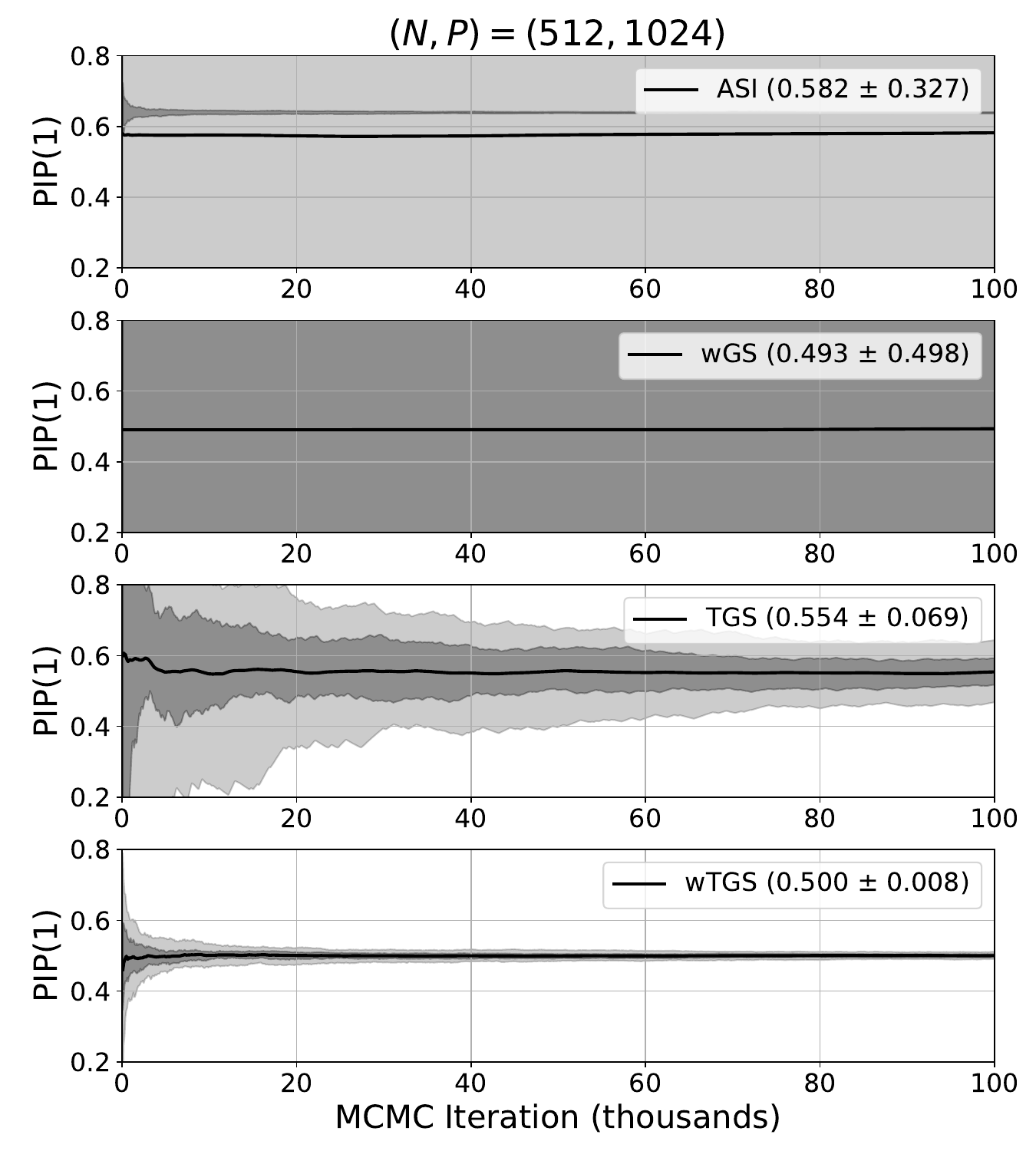}
\label{fig:corr512a}
\end{minipage}
\quad
\begin{minipage}[b]{0.405\linewidth}
    \includegraphics[width=\linewidth]{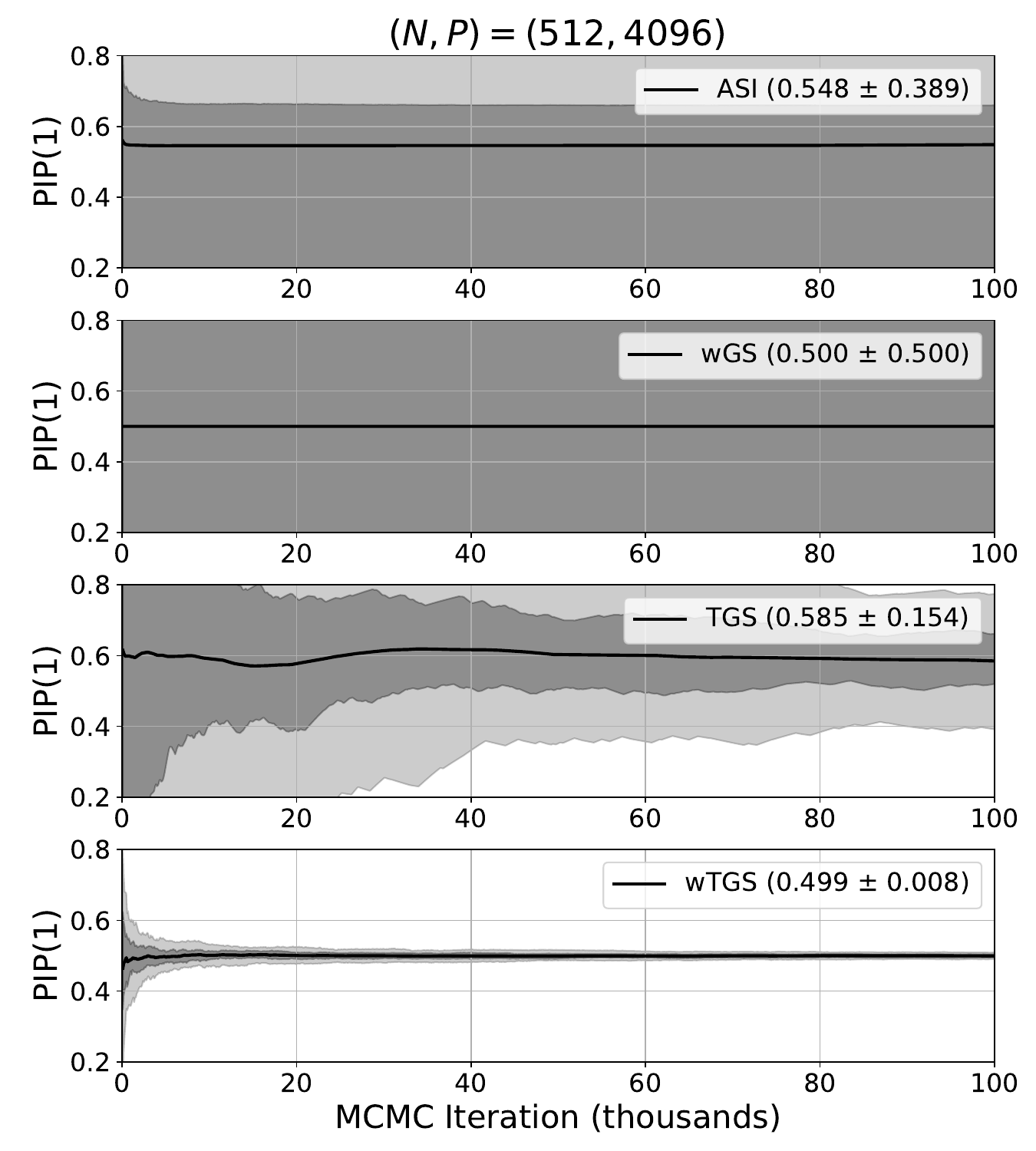}
\label{fig:corr512b}
\end{minipage}
    \caption{This is a companion figure to Fig.~\ref{fig:corr}. We depict posterior inclusion probability (PIP) estimates for the first covariate in the scenario
    described in Sec.~\ref{sec:corr} for four different MCMC methods.
    On the left we depict results for a dataset with $(N, P)=(512, 1024)$, while on the right we depict results
    for a dataset with $(N, P)=(512,4096)$.
    At each iteration $t$ the PIP is computed using all samples obtained through iteration $t$.
    The mean PIP is depicted with a solid black line and light and dark grey confidence intervals
    denote $10\%\!-\!90\%$ and $30\%\!-\!70\%$ quantiles, respectively.
    The true PIP is almost exactly $\tfrac{1}{2}$.
    In each case we run 100 independent chains.
    For each method we also report the final PIP estimate (mean and standard deviation) in parentheses.
    }
\label{fig:tracecorrapp}
\end{figure}

Additional trace plots analogous to Fig.~\ref{fig:corr} in the main text are depicted
in Fig.~\ref{fig:tracecorrapp}.
In Table \ref{table:cancer} we report PIP estimates for top hits in the cancer experiment
in Sec.~\ref{sec:cancer}.

\definecolor{Gray}{gray}{0.80}

\begin{table}[]
 \resizebox{.485\textwidth}{!}{
\begin{tabular}{l|l|l|l}
\rowcolor{Gray} Gene & wTGS-5M & wTGS-250k & ASI-250k \\ \hline
DUSP4 & 1.000 & 1.000 / 1.000 & 1.000 / 1.000 \\ \hline
PPP2R3A & 0.669 & 0.576 / 0.647 & 0.466 / 0.318 \\ \hline
MIA & 0.383 & 0.338 / 0.365 & 0.282 / 0.138 \\ \hline
KRT80 & 0.272 & 0.364 / 0.296 & 0.502 / 0.482 \\ \hline
RELN & 0.243 & 0.286 / 0.219 & 0.346 / 0.502 \\ \hline
ZNF132 & 0.096 & 0.124 / 0.093 & 0.120 / 0.142 \\ \hline
TRIM51 & 0.094 & 0.075 / 0.099 & 0.080 / 0.053 \\ \hline
ZNF471 & 0.083 & 0.107 / 0.081 & 0.139 / 0.181 \\ \hline
S100B & 0.063 & 0.053 / 0.068 & 0.028 / 0.022 \\ \hline
ZNF571 & 0.062 & 0.065 / 0.058 & 0.064 / 0.065 \\ \hline
ZNF304 & 0.060 & 0.067 / 0.069 & 0.095 / 0.068 \\ \hline
ZNF772 & 0.040 & 0.039 / 0.034 & 0.044 / 0.028 \\ \hline
RXRG & 0.032 & 0.026 / 0.025 & 0.010 / 0.028 \\ \hline
ZNF17 & 0.031 & 0.033 / 0.037 & 0.047 / 0.040 \\ \hline
ZNF134 & 0.026 & 0.026 / 0.029 & 0.026 / 0.033 \\ \hline
KRT7 & 0.025 & 0.025 / 0.021 & 0.016 / 0.206 \\ \hline
ZNF71 & 0.020 & 0.022 / 0.014 & 0.045 / 0.023 \\ \hline
CCIN & 0.019 & 0.035 / 0.037 & 0.019 / 0.026 \\ \hline
ZNF419 & 0.018 & 0.019 / 0.017 & 0.018 / 0.006 \\ \hline
ZMYM3 & 0.017 & 0.016 / 0.027 & 0.014 / 0.036 \\ \hline
\end{tabular} 
}
\resizebox{.505\textwidth}{!}{
\begin{tabular}{l|l|l|l}
\rowcolor{Gray} Gene & wTGS-5M & wTGS-250k & ASI-250k \\ \hline
HNF1B & 1.000 & 1.000 / 1.000 & 1.000 / 1.000 \\ \hline
CAP2 & 0.323 & 0.322 / 0.354 & 0.079 / 0.068 \\ \hline
C12orf54 & 0.172 & 0.113 / 0.152 & 0.145 / 0.005 \\ \hline
AQP1 & 0.122 & 0.127 / 0.128 & 0.061 / 0.096 \\ \hline
FAM43B & 0.067 & 0.059 / 0.050 & 0.013 / 0.077 \\ \hline
KLRF1 & 0.063 & 0.059 / 0.065 & 0.032 / 0.034 \\ \hline
ARMC4 & 0.059 & 0.062 / 0.035 & 0.027 / 0.120 \\ \hline
SERPINE1 & 0.050 & 0.047 / 0.052 & 0.040 / 0.003 \\ \hline
CLIC6 & 0.049 & 0.050 / 0.057 & 0.013 / 0.094 \\ \hline
GSDME & 0.048 & 0.056 / 0.050 & 0.101 / 0.066 \\ \hline
UGCG & 0.044 & 0.042 / 0.034 & 0.108 / 0.093 \\ \hline
NEK6 & 0.039 & 0.041 / 0.041 & 0.019 / 0.005 \\ \hline
SERPINA10 & 0.032 & 0.027 / 0.026 & 0.007 / 0.017 \\ \hline
ECH1 & 0.029 & 0.028 / 0.022 & 0.054 / 0.029 \\ \hline
KIF1C & 0.029 & 0.034 / 0.024 & 0.066 / 0.081 \\ \hline
S100A4 & 0.028 & 0.029 / 0.025 & 0.083 / 0.009 \\ \hline
MSANTD3 & 0.023 & 0.029 / 0.023 & 0.005 / 0.006 \\ \hline
PLIN3 & 0.023 & 0.019 / 0.020 & 0.043 / 0.007 \\ \hline
IL4R & 0.021 & 0.018 / 0.021 & 0.016 / 0.003 \\ \hline
SHBG & 0.020 & 0.016 / 0.022 & 0.005 / 0.027 \\ \hline
\end{tabular}
}
    \vspace{2mm}
\caption{These tables are companions to Fig.~\ref{fig:cancer}-\ref{fig:cancerbig}. 
    We depict PIP estimates for DUSP4 (left) and HNF1B (right). In each case we
    include the result from a wTGS run with five million samples as well as two shorter
    runs from wTGS and ASI. We depict the top 20 genes as determined by the long wTGS run.
    The much lower variance and higher accuracy of wTGS are apparent.}
\label{table:cancer}
\end{table}

\subsection{Experimental details}
\label{app:exp}

Unless specified otherwise we set $\tau = 0.01$ and $h = 5 /P$ throughout our experiments.

\paragraph{wGS/TGS/wTGS}

We choose the exploration parameter $\epsilon$ to be $\epsilon=5$ (see Eqn.~\ref{eqn:etadefn}).
We use the $\xi$-adaptation scheme described in Sec.~\ref{app:xi}.

\paragraph{ASI}

ASI has several hyperparameters which we set as follows.
We set the exponent $\lambda_{\rm ASI}$ that controls adaptation to $\lambda_{\rm ASI} = 0.75$.
We set $\epsilon_{\rm ASI}=0.1/P$ as suggested by the authors. We target an acceptance probability of $\tau_{\rm ASI}=0.25$.

\paragraph{Runtime experiment}
For each value of $N$ and $P$ we run each MCMC chain for 2000 burn-in iterations and report iteration times averaged
over a subsequent $10^4$ iterations.

\paragraph{Correlated covariates scenario}

The covariates for $p=3,4,...,P$ are generated independently from a standard Normal distribution:
$X_{n,p} \sim \NN(0, 1)$ for all $n$. We then generate $z \in \RR^N$ with $z_n\sim \NN(0, 1)$ and
set $X_{n, p=1} \sim \NN(z_n, 10^{-4})$ and  $X_{n, p=2} \sim \NN(z_n, 10^{-4})$. That is the first
two covariates are almost identical apart from a small amount of noise. 
We then generate the responses $Y_n$ using success logits given by $\psi_n = z_n$. 
The total count $\TC_n$ for each data point is set to 10.
Consequently the true posterior concentrates on two modes with $\gamma = (1, 0, 0, ...)$ and  $\gamma = (0, 1, 0, ...)$. 
We set $h=1/P$ and run each algorithm for 10 thousand burn-in/warmup iterations and use the subsequent 100 thousand samples for analysis.

\paragraph{Cancer data}

All chains are run for 25 thousand burn-in/warmup iterations.

\paragraph{Hospital data}

We run wTGS for 10 thousand burn-in iterations and use the subsequent 100 thousand samples for analysis.
The $899$ held-out patients are chosen at random.
We use a random walk proposal scale for $\log \nu$ of $0.03$.
We set $\psi_0$ to be the logarithm of the mean of the observed $Y$ (this is equivalent to shifting the prior
mean of the bias $\beta_0$; see Sec~\ref{app:nb}).

\paragraph{Health survey data}

We run wTGS for 10 thousand burn-in iterations and use the subsequent 100 thousand samples for analysis.
We use a random walk proposal scale for $\log \nu$ of $0.03$.
We set $\psi_0$ to be the logarithm of the mean of the observed $Y$ (this is equivalent to shifting the prior
mean of the bias $\beta_0$; see Sec~\ref{app:nb}).

\end{document}